\documentclass[showkeys,nofootinbib,prd]{revtex4}

\usepackage{amsmath}
\usepackage{amsfonts}
\usepackage{amssymb}
\usepackage{amsthm}
\usepackage{mathtools}
\usepackage{subfigure}
\DeclareFontFamily{U}{mathb}{\hyphenchar\font45}
\DeclareFontShape{U}{mathb}{m}{n}{
      <5> <6> <7> <8> <9> <10> gen * mathb
      <10.95> mathb10 <12> <14.4> <17.28> <20.74> <24.88> mathb12
      }{}
\DeclareSymbolFont{mathb}{U}{mathb}{m}{n}

\DeclareMathSymbol{\Sun}{3}{mathb}{"40}
\newcommand\omicron{o}
\allowdisplaybreaks

\begin{document}
\title{Quasi-normal modes of a multi-dimensional rotating Kerr black hole}
\author{Mattia Villani}
\affiliation{University of Urbino Carlo Bo, Department of Pure and Applied Sciences (DiSPeA), Via Santa Chiara, 27, Urbino (PU), 61029, Italy}
\email{mattia.villani@uniurb.it}

\begin{abstract}
The aim of this paper is to present a general way to calculate quasi-normal modes (QNM) of the Teukolsky equation for higher dimensional ($d>4$) Kerr spacetime with compactified extra dimensions. In order to do so, we develop a formalism derived from spinors: we call it multispinor formalism. It is based on vectors of two-spinors and permits us to develop a formalism analogous to that of Newman-Penrose in 4d. From this we show how to derive the Teukolsky equation for gravitational perturbations and calculate the QNM. In order to keep calculations simple we fix, as an example, the dimension number to be six, but the work can be readily generalized to other spacetime dimensions.
\end{abstract}
\maketitle

\section{Introduction}
Spinors were first introduced in geometry by \'Elie Cartan in 1913 \cite{sp1} and in the twenties of XX century, during the development of Quantum Mechanics, it was discovered their importance in physics (and also given their current name). Spinors give a linear representation of the rotation group in any dimension $d$, {each spinor having $2^{\lfloor d/2 \rfloor}$ components \cite{EC,sup}}. They also give a representation of the Lorentz and Poincar\'e groups (see, for example, \cite{supc,RP}). Spinors also have a fundamental role in quantum field theory, where they describe fermions, as was discovered by Dirac (see, for example, \cite{SW1,SW2} and references therein).

The role of spinors in 4d General Relativity (GR) was elucidated by several authors in the 1950s and 1960s see, for example, \cite{bade,P1,NP,witten,RP,RP2} and references therein. In particular, it was shown how to rewrite the tensor formulation of GR in terms of spinors and it was found a particularly convenient formalism, the Newman-Penrose formalism \cite{NP} which has several applications in GR; among them we quote: the study, classification and derivation of spacetime metrics in GR (see \cite{P1,RP2}, the account \cite{stephani} and references therein) and the development of Teukolsky's equations for the perturbation of rotating Kerr black holes \cite{teu1,teu2} and all its application, see for example \cite{lrr,kono}.

In the paper \cite{hd}, authors study quasi-normal modes (QNM) of a d-dimensional regular Schwarzschild black hole. {QNM for higher dimensional Kerr black hole were calculated in \cite{highKerr1,highKerr2,highKerr3}, for example. In \cite{highKerr4} a discussion on the detectability of these QNM and of compactified dimensions is presented.} In this paper, we aim at presenting a general way to calculate the QNM of a rotating d-dimensional Kerr black holes with the additional dimensions compactified to a hypersphere, by deriving the higher dimensional analog of the Teukolsky equation. In order to do so, we develop a formalism based on vectors of two-spinors (we call them multispinors) and then an higher dimensional analog of the Newman-Penrose formalism, deriving, among the other things, the Bianchi identities; finally, we derive the Teukolsky equation for gravitational perturbations of rotating Kerr black hole and calculate the QNM. At the beginning of this paper, we shall be as general as possible, almost never specifying the number of dimensions of the spacetime. In later Sections, however, we shall consider, for definiteness and simplicity, a 6d spacetime. 

We are aware of the fact that the higher dimensional Teukolsky equation is not separable for generic spacetime, but since theories that predict higher dimensional spacetime also predict that the additional dimensions are compactified, we shall only consider this case and we shall find that the Teukolsky equation is separable.

The formalism presented here is not applicable only to the Teukolsky equation; in a forthcoming paper, we shall show how the classification of higher dimensional spacetimes presented in \cite{hdc1,hdc2,hdc3} can be easily obtained also with multispinors along the lines of what was done  in \cite{P1,RP2}; we also notice that the Bianchi identities calculated here are treated more easily with multispinors than with the formalism of \cite{hdc4}. Moreover, in another paper, we shall show how to apply the multispinor formalism to the derivation and characterization of higher dimensional spacetime metrics similarly to the account \cite{stephani} by first proving the higher dimensional analog of the Goldberg-Sachs theorem. {Moreover, once a proper vielbein has been selected, one can apply our method also to the study of rotating black holes in different theories of gravity or in the presence of dark matter, see for example \cite{ref1,ref2,ref3}.}

The plan for this paper is as follows: in Sections \ref{sec:multi} and \ref{sec:geom}, we shall introduce multispinors and give a geometrical interpretation; in Section \ref{sec:lor}, we shall show how with multispinors it is possible to derive the Lorentz transformation; in Section \ref{sec:basics}, we shall discuss the basics of multispinor formalism, like the spin frames and how to rewrite tensors; in Section \ref{sec:curv} we shall introduce the curvature multispinors and the Weyl and Ricci scalars; in Section \ref{sec:bia} we derive the higher dimensional Bianchi identities. In Section \ref{sec:NP} we discuss the Newman-Penrose formalism for 6d metrics. In Section \ref{sec:teueq} we derive the 6d Teukolsky equation for gravitational perturbations and in Section \ref{sec:QNM} we calculate the QNM. Finally, in Section \ref{sec:concl}, we conclude our presentation. There are also two appendices, in which we report the full expressions of the derivatives of the spin coefficients and of the Bianchi identities.

{A few words on notation: an overbar indicates complex conjugation; starting from Section \ref{sec:teueq} the symbol $\Delta$ has the usual meaning in Kerr metric and is not the Newman-Penrose derivative anymore (we didn't feel right to change a well established notation); moreover we suppress all the $G$ and $c$ setting them to 1.}

\section{Multidimensional spacetimes}
\label{sec:multi}
We start by considering a generic 6d spacetime with coordinates $\{t,x_0,\dots,x_4\}$, where $t$ is the time coordinate and $x_i$ are the spatial coordinates. We follow \cite{EC,RP} and introduce the matrix:
\begin{equation}\label{eq:mat1}
X_6= \left( \begin{array}{cccc}
t+x_0 & x_1+i x_2 & x_4+i x_3 & 0\\
x_1-i x_2 & t-x_0 &0 & x_4+i x_3\\
x_4-i x_3 &0 & t-x_0 & -x_1-i x_2\\
0 & x_4-i x_3 & -x_1+i x_2 & t+x_0
\end{array} \right);
\end{equation}
{this matrix has determinant ${(t^2-x_0^2-x_1^2-x_2^2-x_3^2-x_4^2-x_5^2)^2}$, the line element squared.} We notice that this matrix can be written as follows:
\begin{equation}
X_6= t\,(\boldsymbol{1}_2\otimes \boldsymbol{1}_2)+x_0\,(\sigma_3\otimes\sigma_3)+x_1\, (\sigma_3\otimes\sigma_1)-x_2\,(\sigma_3\otimes\sigma_2)- x_3\,(\sigma_2\otimes \boldsymbol{1}_2)+x_4\,(\sigma_1\otimes \boldsymbol{1}_2) 
\end{equation}
where $\boldsymbol{1}_2$ is the 2d identity matrix and $\sigma_i$ are the Pauli matrices. So $X_6$ is a linear combination of tensor products of $SU(2)$ matrices.

We now consider two two-spinors $\boldsymbol{\kappa}_1=\{\xi_1,\xi_2\}$ and $\boldsymbol{\kappa}_2=\{\xi_3,\xi_4\}$ and consider the matrix given by:
\begin{equation}
M=2(\sigma_3\otimes(\boldsymbol{\kappa}_1\overline{\boldsymbol{\kappa}}_1^T))+2((\boldsymbol{\kappa}_2\overline{\boldsymbol{\kappa}}_2^T)\otimes \boldsymbol{1}_2)
\end{equation}
where an overline indicates the complex conjugate and ${}^T$ indicates a transpose. Explicitly, $M$ is given by:
\begin{equation}\label{eq:spin1}
M=\left( \begin{array}{cccc}
\xi_1\overline{\xi}_1+\xi_3\overline{\xi}_3 & \xi_1 \overline{\xi}_2 & \xi_3\overline{\xi}_3 &0\\
\xi_2\overline{\xi}_1 & \xi_2\overline{\xi}_2+\xi_3\overline{\xi}_3 & 0 & \xi_3\overline{\xi}_3\\
\xi_4\overline{\xi}_3 & 0 & -\xi_1\overline{\xi}_1+\xi_4\overline{\xi}_4 & - \xi_1\overline{\xi}_2\\
0 & \xi_4\overline{\xi}_3 & -\xi_2\overline{\xi}_1 & -\xi_2\overline{\xi}_2+\xi_4\overline{\xi}_4\\
\end{array} \right).
\end{equation}

By comparing $X_6$ and $M$ we can find an expression for the coordinates $x_i$ and $t$ in terms of the spinors components:
\begin{align}
x_0 &= \xi_1\overline{\xi}_1-\xi_2\overline{\xi}_2,\\
x_1 &= \xi_2\overline{\xi}_1+\xi_1\overline{\xi}_2,\\
x_2 &= i(\xi_2\overline{\xi}_1-\xi_1\overline{\xi}_2),\\
x_3 &= i(\xi_4\overline{\xi}_3-\xi_3\overline{\xi}_4),\\
x_4 &= \xi_4\overline{\xi}_3+\xi_3\overline{\xi}_4,\\
t   &= 2\xi_4\overline{\xi}_4-\xi_1\overline{\xi}_1-\xi_2\overline{\xi}_2,\\
    &= 2\xi_3\overline{\xi}_3+\xi_1\overline{\xi}_1+\xi_2\overline{\xi}_2.
\end{align}
From the last two expressions for $t$, we derive the consistency relation:
\begin{equation}\label{eq:consis}
-\xi_3\overline{\xi}_3+\xi_4\overline{\xi}_4=\xi_1\overline{\xi}_1+\xi_2\overline{\xi}_2,
\end{equation}
which shows that the components of the two spinors are not independent.

The above construction can be iterated for higher dimensional spacetimes. For example, in 8d we have:
\begin{equation}
\begin{split}
X_8&= t\,(\boldsymbol{1}_2\otimes \boldsymbol{1}_2\otimes \boldsymbol{1}_2)+x_0\,(\sigma_3\otimes\sigma_3\otimes\sigma_3)+x_1\, (\sigma_3\otimes\sigma_3\otimes\sigma_1)-x_2\,(\sigma_3\otimes\sigma_3\otimes\sigma_2)+\\
&- x_3\,(\sigma_3\otimes\sigma_2\otimes \boldsymbol{1}_2)+x_4\,(\sigma_3\otimes\sigma_1\otimes \boldsymbol{1}_2) -x_5 \, (\boldsymbol{1}_2\otimes \boldsymbol{1}_2 \otimes \sigma_2)+ x_6\, (\boldsymbol{1}_2\otimes \boldsymbol{1}_2 \otimes \sigma_1).
\end{split}
\end{equation}
We now need to introduce three two-spinors: $\boldsymbol{\kappa}_1=\{\xi_1,\xi_2\}$, $\boldsymbol{\kappa}_2=\{\xi_3,\xi_4\}$ and $\boldsymbol{\kappa}_3=\{\xi_5,\xi_6\}$ and consider the matrix:
\begin{equation}
M=2(\sigma_3\otimes\sigma_3\otimes(\boldsymbol{\kappa}_1\overline{\boldsymbol{\kappa}}_1^T))+2(\sigma_3\otimes(\boldsymbol{\kappa}_2\overline{\boldsymbol{\kappa}}_2^T)\otimes \boldsymbol{1}_2)+2((\boldsymbol{\kappa}_3\overline{\boldsymbol{\kappa}}_3^T)\otimes \boldsymbol{1}_2\otimes \boldsymbol{1}_2).
\end{equation}
In the same way as above, we can find relations between the spinors components and the coordinates $t$ and $x_i$ and also the two consistency relations:
\begin{align}
\xi_1\overline{\xi}_1+\xi_2\overline{\xi}_2&=-\xi_3\overline{\xi}_3+\xi_4\overline{\xi}_4,\\
\xi_3\overline{\xi}_3+\xi_4\overline{\xi}_4&=-\xi_5\overline{\xi}_5+\xi_6\overline{\xi}_6.
\end{align}
In 10d, we shall need four two-spinors and so on.

The above construction can also be repeated for odd-dimensional spacetime just by dropping $x_0$, as discussed in \cite{EC}. It also works with Riemannian spaces of any dimension, one just has to drop the coordinate $t$. We notice that in the Riemannian case there will be no consistency relation and the components of the spinors will be independent.

\section{Geometrical interpretation of the construction}
\label{sec:geom}

As described in \cite{RP}, given a Minkowski spacetime, one is often concerned with the null directions. In 4d, these directions constitute two abstract spheres $\mathcal{S}^\pm$ corresponding to the future (+) and past (--) directions. Given a reference frame in a Minkowski space and the null-cone structure, one can impose the correspondence between the spheres $S^\pm$ of the future (+) and past (--) light cone and $\mathcal{S}^\pm$, so that these spheres become the (future and past) celestial spheres. The spheres $S^\pm$ are obtained by intersecting the future and past light-cone with infinite planes passing through  $t=\pm 1$.\footnote{One chooses $t=\pm 1$, because in this way $S^\pm$ are unit spheres, since in terms of Minkowski coordinates, one has (on the null-cone) $1=x^2+y^2+z^2$.} The inner part of $\mathcal{S}^\pm$ constitutes time-like directions contained inside the null-cone.

This construction can be extended to higher dimensional spacetime with $d>4$, but now the celestial sphere will become $d-2$ dimensional hyper-spheres $\mathcal{S}^\pm_{d-2}$.

In \cite{RP}, it is shown how to derive the correspondence between null directions and spinors in $S^\pm$. In our treatment, we are considering two-spheres sections of  $\mathcal{S}^\pm_{d-2}$ by fixing the coordinates $x_0$ and $t$ and considering in turn pairs of the other coordinates. In each of these spheres one can repeat the construction of \cite{RP} and obtain the correspondence between null coordinates and spinors. Since the $x_0$ coordinate is common to all spheres, the derived expressions for $x_0$ in terms of spinors must all be consistent; the same is true for the time coordinate $t$, which is also fixed by the chosen reference frame: this is the source of the consistency relations like \eqref{eq:consis}.

\section{Lorentz transformations}
\label{sec:lor}
In the 4d case, a Lorentz transformation can be written using a $SU(2)$ operator $\mathbf{A}$ as follows \cite{RP}:
\begin{equation}
\mathbf{A}\cdot\left( \begin{array}{cc}
t+z & x+iy\\
x-iy & t-z
\end{array} \right)\cdot\overline{\mathbf{A}}^T
\end{equation}
Spinors $\boldsymbol{\kappa}$ transform with $\mathbf{A}$ as follows: 
\begin{equation}\label{eq:transf}
\boldsymbol{\kappa}^\prime=\boldsymbol{A}\cdot\boldsymbol{\kappa}.
\end{equation}

It is natural to try to use $SU(2)$ operators also in higher dimensions. We now focus on 6d and consider two $SU(2)$ operators:
\begin{equation}\label{eq:matrices}
\mathbf{A}=\left( \begin{array}{cc}
a_1 & b_1\\
c_1 & d_1
\end{array} \right) \qquad \mathbf{B}=\left( \begin{array}{cc}
a_2 & b_2\\
c_2 & d_2
\end{array} \right) \qquad a_id_i-c_ib_i = 1.
\end{equation}
and write the matrix:
\begin{equation}
\mathbf{T}= \dfrac{1}{\sqrt{2}}\,(\sigma_3\otimes\mathbf{A})+\dfrac{1}{\sqrt{2}}\,(\mathbf{B}\otimes \boldsymbol{1}_2)
\end{equation}
It can be checked that it has $\det(\mathbf{T})=1$ if
\begin{equation}\label{eq:condition}
a_1+d_1=-a_2+d_2.
\end{equation}
So the expression $\mathbf{T}\cdot X_6\cdot \overline{\mathbf{T}}^T$ indeed leaves the determinant of $X_6$ invariant{, and with it the line element}. The two-spinor $\boldsymbol{\kappa}_1$ introduced above in the 6d case transforms with the matrix $\boldsymbol{A}$, while $\boldsymbol{\kappa}_2$ trasforms with $\boldsymbol{B}$, similarly to equation \eqref{eq:transf}.

The 8d case is similar: there are now 3 $SU(2)$ matrices:
\begin{equation}
\mathbf{A}=\left( \begin{array}{cc}
a_1 & b_1\\
c_1 & d_1
\end{array} \right) \qquad \mathbf{B}=\left( \begin{array}{cc}
a_2 & b_2\\
c_2 & d_2
\end{array} \right) \qquad \mathbf{C}=\left( \begin{array}{cc}
a_3 & b_3\\
c_3 & d_3
\end{array} \right) \qquad a_id_i-c_ib_i = 1.
\end{equation}
and the Lorentz transformation is given by:
\begin{equation}
\mathbf{T}= \dfrac{1}{\sqrt{3}}\,(\sigma_3\otimes\sigma_3\otimes\mathbf{A})+\dfrac{1}{\sqrt{3}}\,(\sigma_3\otimes\mathbf{B}\otimes \boldsymbol{1}_2)+\dfrac{1}{\sqrt{3}}\,(\mathbf{C}\otimes \boldsymbol{1}_2\otimes \boldsymbol{1}_2)
\end{equation}
with the conditions
\begin{align}
a_1+d_1 &=-a_2+d_2\\
a_2+d_2 &=-a_3+d_3
\end{align}
We notice that this completely fixes the two diagonal elements of $\mathbf{B}$. Since it also has unit determinant, $\mathbf{B}$ has only one independent component. The three spinors $\boldsymbol{\kappa}_1$, $\boldsymbol{\kappa}_2$ and $\boldsymbol{\kappa}_3$ transform respectively, with $\boldsymbol{A}$, $\boldsymbol{B}$ and $\boldsymbol{C}$, as above.

This construction can be iterated to higher dimension in the trivial way. This also works in odd dimensional spacetimes.

We now consider  6d spacetimes and try to write $\boldsymbol{A}$ and $\boldsymbol{B}$ for particular Lorentz tranformations.

\subsection{Rotation around $x_0$}

It can be checked that a rotation around $x_0$ can be written as:
\begin{equation}
R_0=\left( \begin{array}{cccc}
\exp(i\,\theta_0/2) & 0 & 0 & 0  \\
0 & \exp(-i\,\theta_0/2) & 0 & 0 \\
0 & 0 & -\exp(-i\,\theta_0/2) & 0\\
0 & 0 & 0 & -\exp(i\,\theta_0/2)
\end{array} \right).
\end{equation}
Then, if we introduce the two matrices:
\begin{equation}
\boldsymbol{A} = \left( \begin{array}{cc}
a_1\, \exp(i\,\theta_0/2) + a_2\, \exp(-i\,\theta_0/2) & 0\\
0 & b_1\, \exp(-i\,\theta_0/2) + b_2\, \exp(i\,\theta_0/2)
\end{array} \right),
\end{equation}
\begin{equation}
\boldsymbol{B} = \left( \begin{array}{cc}
c_1\, \exp(i\,\theta_0/2) + c_2\, \exp(-i\,\theta_0/2) & 0\\
0 & d_2\, \exp(-i\,\theta_0/2) + d_1\, \exp(i\,\theta_0/2) 
\end{array} \right)
\end{equation}
and the matrix $\boldsymbol{T}$ as above, we find that we must have (imposing also that the determinant is unitary):
\begin{equation}
\boldsymbol{A} = \left( \begin{array}{cc}
-i+\exp(i\,\theta_0/2) & 0\\
0 & -i+\exp(-i\,\theta_0/2)
\end{array} \right), 
\end{equation}
\begin{equation}
\boldsymbol{B} = \left( \begin{array}{cc}
i & 0 \\
0 & -i
\end{array} \right).
\end{equation}
These are the matrices that define a rotation by $\theta_0$ around the coordinate $x_0$.

\subsection{Rotation around $x_2$}
A rotation around $x_2$ can be written as:
\begin{equation}
R_2=\left( \begin{array}{cccc}
-\cos(\theta_2) & - \dfrac{\sin(\theta_2)}{\sqrt{2}} & - \dfrac{\sin(\theta_2)}{\sqrt{2}} & 0 \\
\dfrac{\sin(\theta_2)}{\sqrt{2}} & -\cos(\theta_2) & 0 & - \dfrac{\sin(\theta_2)}{\sqrt{2}}\\
- \dfrac{\sin(\theta_2)}{\sqrt{2}} & 0 & - \cos(\theta_2) & \dfrac{\sin(\theta_2)}{\sqrt{2}} \\
0 & - \dfrac{\sin(\theta_2)}{\sqrt{2}} & - \dfrac{\sin(\theta_2)}{\sqrt{2}} & - \cos(\theta_2) 
\end{array} \right).
\end{equation}

If we introduce the two matrices
\begin{equation}
\boldsymbol{A} = \left( \begin{array}{cc}
d \, \cos(\theta_2) & -a\,\sin(\theta_2) \\
a\, \sin(\theta_2) & d\,\cos(\theta_2)
\end{array} \right),
\end{equation}
\begin{equation}
\boldsymbol{B} = \left( \begin{array}{cc}
e \, \cos(\theta_2) & i\,b\,\sin(\theta_2) \\
i\,b\, \sin(\theta_2) & e\,\cos(\theta_2)
\end{array} \right),
\end{equation}
Using again the matrix $\boldsymbol{T}$ defined above and the fact that the determinant must be unitary, we can reproduce $R_2$ if we define:
\begin{equation}
a=\dfrac{1}{\sqrt{2}}, \quad b=\dfrac{i}{\sqrt{2}}, \quad d=\dfrac{1}{\sqrt{2}}, \quad e=1-\dfrac{1}{\sqrt{2}}.
\end{equation}

\subsection{Boost in the $x_0$ direction}
It can be checked that a boost in the $x_0$ direction can be obtained with the matrix:
\begin{equation}
B_0=\left( \begin{array}{cccc}
w & 0 & 0 & 0\\
0 & \dfrac{1}{\sqrt{w}} & 0 & 0\\
0 & 0 & w & 0\\
0 & 0 & 0 & \dfrac{1}{\sqrt{w}}
\end{array} \right)
\end{equation}
where $w$ is the relativistic Doppler factor ($\ln(w)=\tanh(v)$ is the rapidity), defined as
\begin{equation}
w=\sqrt{\dfrac{1+v}{1-v}},
\end{equation}
where $v$ is the velocity, see \cite{RP}. We introduce the matrices:
\begin{equation}
\boldsymbol{A}=2\,\left( \begin{array}{cc}
a & e\\
e & b
\end{array} \right), \quad \boldsymbol{B}=2\,\left( \begin{array}{cc}
c & f\\
f & d
\end{array} \right).
\end{equation}
Using the matrix $\boldsymbol{T}$ defined above, we can reproduce $B_0$ if we impose:
\begin{equation}
\begin{split}
&a=\dfrac{1}{2}\, \left( 1-w-\sqrt{1+w^2} \right), \quad b=\dfrac{-1+w-\sqrt{1+w^2}}{2w},\\ &c=\dfrac{1}{2}\, \left( -1-w+\sqrt{1+w^2} \right), \quad d=-\dfrac{1+w+\sqrt{1+w^2}}{2w}, \quad e=f=0.
\end{split}
\end{equation}

\subsection{The identity and the composition law}
The identity is given by:
\begin{equation}
\boldsymbol{I}=\sqrt{2}\,\boldsymbol{1}_2\otimes \boldsymbol{1}_2.
\end{equation}

In order to find the composition law we define:
\begin{equation}
\boldsymbol{A}=\dfrac{1}{\sqrt{2}}\, \Big( \sigma_3\otimes \boldsymbol{A}_1 + \boldsymbol{A}_2\otimes \boldsymbol{1}_2 \Big), \quad \boldsymbol{B}=\dfrac{1}{\sqrt{2}}\, \Big( \sigma_3\otimes \boldsymbol{B}_1 + \boldsymbol{B}_2\otimes \boldsymbol{1}_2 \Big).
\end{equation}
The composition $\boldsymbol{B}\cdot\boldsymbol{A}$ is defined as follows, as can be checked by expanding the products:
\begin{equation}
\boldsymbol{B}\cdot\boldsymbol{A}=\dfrac{1}{2}\, \Big[ \boldsymbol{1}_2\otimes(\boldsymbol{B}_1\cdot\boldsymbol{A}_1)+(\boldsymbol{B}_2\cdot\boldsymbol{A}_2)\otimes\boldsymbol{1}_2+(\sigma_3\cdot\boldsymbol{A}_2)\otimes(\boldsymbol{B}_1\cdot\boldsymbol{1}_2)+(\boldsymbol{B}_2\cdot\sigma_3)\otimes(\boldsymbol{1}_2\cdot\boldsymbol{A}_1) \Big],
\end{equation}
where we have used $\sigma_3\cdot\sigma_3=\boldsymbol{1}_2=\boldsymbol{1}_2\cdot\boldsymbol{1}_2$. If the the matrices $\boldsymbol{A}_i$ and $\boldsymbol{B}_i$ ($i=\{1,2\}$) satisfy the condition \eqref{eq:condition} the above matrix automatically has unit determinant, as it should. With this we have proven that the matrices $\boldsymbol{T}$ form a group, which is equivalent to the Lorentz group.

\section{The basics of multispinor formalism}
\label{sec:basics}

\subsection{Multispinor scalar product}
In 4d, given the two spinors $\boldsymbol{\alpha}$ and $\boldsymbol{\beta}$, the scalar product is given by, see \cite{RP}:
\begin{equation}
\langle\boldsymbol{\alpha}|\boldsymbol{\beta}\rangle = \overline{\alpha}^1\beta_1-\overline{\alpha}^2\beta_2,
\end{equation}
where $\alpha^i$ and $\beta^i$ ($i=\{1,2\}$) are the components of the spinors. We can define an antisymmetric $\boldsymbol{\epsilon}_{AB}=-\boldsymbol{\epsilon}_{BA}$ spinor, such that:
\begin{equation}
\langle\boldsymbol{\alpha}|\boldsymbol{\beta}\rangle=\boldsymbol{\epsilon}_{AB}\overline{\alpha}^A\beta^B.
\end{equation}
This antisymmetric spinor can be used to raise and lower spinor indices:
\begin{equation}
\boldsymbol{\epsilon}^{AB}\,\kappa_A=\kappa^B.
\end{equation}
Moreover, we have $\boldsymbol{\epsilon}^{AB}\boldsymbol{\epsilon}_{AC}=\delta^B_C$, where $\delta_{AB}$ is the Kronecker delta.

We can define an analogous operation for multispinors. For definiteness we consider the 6d case, but the equations can be easily extended to higher dimensions.

We introduce the two multispinors $[\boldsymbol{\alpha}_1,\boldsymbol{\alpha}_2]$ and $[\boldsymbol{\beta}_1,\boldsymbol{\beta}_2]$. Then, the scalar product is defined as follows:
\begin{equation}\label{eq:scalar}
\langle\langle[\boldsymbol{\alpha}_1,\boldsymbol{\alpha}_2]|[\boldsymbol{\beta}_1,\boldsymbol{\beta}_2]\rangle\rangle= \langle\boldsymbol{\alpha}_1|\boldsymbol{\beta}_1\rangle\langle\boldsymbol{\alpha}_2|\boldsymbol{\beta}_2\rangle.
\end{equation}
In this case, we need to define an $\boldsymbol{\epsilon}_{AB}$ spinor for each copy of $SU(2)$, so in 6d we have two different antisymmetric spinors. Each of these spinors will act on one, and only one of spinor components of the multispinors. The above construction can be repeated also for higher dimensional spacetimes: there will be correspondingly more factors in the product \eqref{eq:scalar}.

\subsection{Spin frames}

In 4d (see for example \cite{RP,NP}), we can introduce the spin frames, i.e. spinors $\boldsymbol{\iota}$ and $\boldsymbol{\omicron}$ such that:
\begin{equation}
\langle\boldsymbol{\iota}|\boldsymbol{\omicron}\rangle=-1, \qquad \langle\boldsymbol{\iota}|\boldsymbol{\iota}\rangle=\langle\boldsymbol{\omicron}|\boldsymbol{\omicron}\rangle=0.
\end{equation}
In this way, we can define the components of a spinor $\boldsymbol{\kappa}$ as follows:
\begin{equation}
\kappa^0=\langle\boldsymbol{\kappa}|\boldsymbol{\iota}\rangle, \quad \kappa^1=-\langle\boldsymbol{\kappa}|\boldsymbol{\omicron}\rangle
\end{equation}
so that $\boldsymbol{\kappa}=\kappa^0\boldsymbol{\omicron}+\kappa^1\boldsymbol{\iota}$ (see \cite{RP}). We have \cite{RP}:
\begin{align}
&\boldsymbol{\epsilon}_{0A}=\boldsymbol{\epsilon}_{A0}=\boldsymbol{\omicron}_A, \quad \boldsymbol{\epsilon}_{1A}=\boldsymbol{\epsilon}_{A1}=\boldsymbol{\iota}_A,\\
&\boldsymbol{\epsilon}^{0A}=\boldsymbol{\epsilon}^{A0}=\boldsymbol{\iota}^A, \quad \boldsymbol{\epsilon}^{1A}=\boldsymbol{\epsilon}^{A1}=\boldsymbol{\omicron}^A
\end{align}

It is convienent to have a single symbol for the spin frame $\boldsymbol{\zeta}_i^{\phantom{k}A}$ (see \cite{NP}), so that:
\begin{equation}
\boldsymbol{\zeta}_0^{\phantom{0}A}=\boldsymbol{\omicron}^A, \quad \boldsymbol{\zeta}_1^{\phantom{0}A}=\boldsymbol{\iota}^A, \quad \overline{\boldsymbol{\zeta}}_{\dot{0}}^{\phantom{0}\dot{A}}=\overline{\boldsymbol{\omicron}}^{\dot{A}}, \quad \overline{\boldsymbol{\zeta}}_{\dot{1}}^{\phantom{0}\dot{A}}=\overline{\boldsymbol{\iota}}^{\dot{A}}.
\end{equation}

For multispinors, we define (in 6d for definiteness): $[\boldsymbol{\iota}_1,\boldsymbol{\iota}_2],$ $[\boldsymbol{\iota}_1,\boldsymbol{\omicron}_2],$ $[\boldsymbol{\omicron}_1,\boldsymbol{\iota}_2]$ and $[\boldsymbol{\omicron}_1,\boldsymbol{\omicron}_2]$, which satisfy:
\begin{align}
\langle\langle[\boldsymbol{\iota}_1,\boldsymbol{\iota}_2],[\boldsymbol{\iota}_1,\boldsymbol{\iota}_2]\rangle\rangle &= \langle\boldsymbol{\iota}_1,\boldsymbol{\iota}_1\rangle\langle\boldsymbol{\iota}_2,\boldsymbol{\iota}_2\rangle=0,\\
\langle\langle[\boldsymbol{\omicron}_1,\boldsymbol{\omicron}_2],[\boldsymbol{\omicron}_1,\boldsymbol{\omicron}_2]\rangle\rangle &= \langle\boldsymbol{\omicron}_1,\boldsymbol{\omicron}_1\rangle\langle\boldsymbol{\omicron}_2,\boldsymbol{\omicron}_2\rangle=0,\\
\langle\langle[\boldsymbol{\iota}_1,\boldsymbol{\iota}_2],[\boldsymbol{\omicron}_1,\boldsymbol{\omicron}_2]\rangle\rangle &= \langle\boldsymbol{\iota}_1,\boldsymbol{\omicron}_1\rangle\langle\boldsymbol{\iota}_2,\boldsymbol{\omicron}_2\rangle=1,\\
\langle\langle[\boldsymbol{\iota}_1,\boldsymbol{\omicron}_2],[\boldsymbol{\omicron}_1,\boldsymbol{\iota}_2]\rangle\rangle &= \langle\boldsymbol{\iota}_1,\boldsymbol{\omicron}_1\rangle\langle\boldsymbol{\omicron}_2,\boldsymbol{\iota}_2\rangle=-1,\\
\langle\langle[\boldsymbol{\omicron}_1,\boldsymbol{\iota}_2],[\boldsymbol{\iota}_1,\boldsymbol{\omicron}_2]\rangle\rangle &= \langle\boldsymbol{\omicron}_1,\boldsymbol{\iota}_1\rangle\langle\boldsymbol{\iota}_2,\boldsymbol{\omicron}_2\rangle=-1,\\
\langle\langle[\boldsymbol{\omicron}_1,\boldsymbol{\omicron}_2],[\boldsymbol{\iota}_1,\boldsymbol{\iota}_2]\rangle &= \langle\boldsymbol{\omicron}_1,\boldsymbol{\iota}_1\rangle\langle\boldsymbol{\omicron}_2,\boldsymbol{\iota}_2\rangle=1.
\end{align}
We also impose $\langle\boldsymbol{\iota}_1,\boldsymbol{\omicron}_2\rangle =\langle\boldsymbol{\iota}_2,\boldsymbol{\omicron}_1\rangle=\langle\boldsymbol{\iota}_2,\boldsymbol{\iota}_1\rangle=\langle\boldsymbol{\omicron}_2,\boldsymbol{\omicron}_1\rangle= 0$.

As above, we define a single symbol for the spin frame $\boldsymbol{\zeta}_{ab}^{\phantom{ab}AB}$:
\begin{equation}
\boldsymbol{\zeta}_{00}^{\phantom{00}AB} =[\boldsymbol{\iota}^A_1,\boldsymbol{\iota}^B_2], \qquad \boldsymbol{\zeta}_{11}^{\phantom{11}AB}=[\boldsymbol{\omicron}^A_1,\boldsymbol{\omicron}^B_2], \qquad \boldsymbol{\zeta}_{10}^{\phantom{10}AB}=[\boldsymbol{\omicron}^A_1,\boldsymbol{\iota}^B_2], \qquad \boldsymbol{\zeta}_{01}^{\phantom{01}AB}=[\boldsymbol{\iota}^A_1,\boldsymbol{\omicron}^B_2],
\end{equation}
and the complex conjugate:
\begin{equation}
\overline{\boldsymbol{\zeta}}_{\dot{0}\dot{0}}^{\phantom{00}AB} =[\overline{\boldsymbol{\iota}}^A_{\dot{1}},\overline{\boldsymbol{\iota}}^B_{\dot{2}}], \qquad \overline{\boldsymbol{\zeta}}_{\dot{1}\dot{1}}^{\phantom{11}AB}=[\overline{\boldsymbol{\omicron}}^A_{\dot{1}},\overline{\boldsymbol{\omicron}}^B_{\dot{2}}] \qquad \overline{\boldsymbol{\zeta}}_{\dot{1}\dot{0}}^{\phantom{10}AB}=[\overline{\boldsymbol{\omicron}}^A_{\dot{1}},\overline{\boldsymbol{\iota}}^B_{\dot{2}}], \qquad \overline{\boldsymbol{\zeta}}_{\dot{0}\dot{1}}^{\phantom{01}AB}=[\overline{\boldsymbol{\iota}}^A_{\dot{1}},\overline{\boldsymbol{\omicron}}^B_{\dot{2}}].
\end{equation}
These spinors will be identified with small indices, as follows:
\begin{align}
&\zeta^{\phantom{a_1b_1}AB}_{a_1a_2}, \qquad a_1,a_2=\{0,1\},\\
&\overline{\zeta}^{\phantom{a_1b_1}AB}_{\dot{a}_1\dot{a}_2}, \qquad \dot{a}_1,\dot{a}_2=\{0,1\}.
\end{align}

In higher dimensional spacetimes the construction is analogous.

\subsection{Tensors in multispinor formalism}
In usual 4d spacetimes, every tensor index can be written with two spinor indices $A\dot{A}$, for example:
\begin{equation}
X_\mu=X_{A\dot{A}}.
\end{equation}

In our case, we have a pair of indices for every copy of $SU(2)$, so there are four indices in 6d ($A_1A_2\dot{B}_1\dot{B}_2$), where the indices $1,2$ are used to indicate to which copy of $SU(2)$ the index refers to. Thus a generic tensor in 6d will be indicated as follows:
\begin{equation}
T^{\mu}_{\phantom{\mu}\nu\rho}=T^{A_1A_2\dot{B}_1\dot{B}_2}_{\phantom{A_1A_2\dot{B}_1\dot{B}_2}C_1C_2\dot{D}_1\dot{D}_2E_1E_2\dot{F}_1\dot{F}_2}
\end{equation}
For higher dimensional spacetimes there will be as many copies of each index as the number of copies of $SU(2)$, so for example there will be three copies in 8d spacetimes, and so on.

We notice that on each pair of spinor indices acts its particular copy of the $\boldsymbol{\epsilon}$ spinor, and only that. One cannot contract pairs of indices coming from different $SU(2)$ copies.

The proliferation of indices seems unavoidable, therefore we introduce multi-index notation:
\begin{equation}
T^{\mu}_{\phantom{\mu}\nu\rho}=T^{\boldsymbol{A}\dot{\boldsymbol{B}}}_{\phantom{\boldsymbol{A}\dot{\boldsymbol{B}}}\boldsymbol{C}\dot{\boldsymbol{D}}\boldsymbol{E}\dot{\boldsymbol{F}}},
\end{equation}
where a bold index is generically given by $\boldsymbol{A}=A_1A_2A_3\dots$, and analogously for dotted indices. This also permits to be as general as possible.

Following \cite{P1,bade}, we define the spinor $\boldsymbol{\sigma}_\mu^{\phantom{\mu}\boldsymbol{A}\dot{\boldsymbol{A}}}$. It satisfies:
\begin{equation}
\boldsymbol{\sigma}_{\mu\phantom{\boldsymbol{A}}\dot{\boldsymbol{B}}}^{\phantom{\mu}\boldsymbol{A}}\boldsymbol{\sigma}_{\nu}^{\phantom{\mu}\boldsymbol{C}\dot{\boldsymbol{B}}} + \boldsymbol{\sigma}_{\nu\phantom{\boldsymbol{A}}\dot{\boldsymbol{B}}}^{\phantom{\mu}\boldsymbol{A}}\boldsymbol{\sigma}_{\mu}^{\phantom{\mu}\boldsymbol{C}\dot{\boldsymbol{B}}}=g_{\mu\nu} \boldsymbol{\epsilon}^{\boldsymbol{A}\boldsymbol{C}},
\end{equation}
where $g_{\mu\nu}$ is the spacetime metric.

With this spinor, we can write, for example:
\begin{align}
X^\mu_{\phantom{\mu}\nu}&= \boldsymbol{\sigma}^\mu_{\phantom{\mu}\boldsymbol{A}\dot{\boldsymbol{B}}}\, X^{\boldsymbol{A}\dot{\boldsymbol{B}}}_{\phantom{\boldsymbol{A}\dot{\boldsymbol{B}}}\boldsymbol{C}\dot{\boldsymbol{D}}}\, \boldsymbol{\sigma}_\nu^{\phantom{\nu}\boldsymbol{C}\dot{\boldsymbol{D}}},\\
X^{\boldsymbol{A}\dot{\boldsymbol{B}}}_{\phantom{\boldsymbol{A}\dot{\boldsymbol{B}}}\boldsymbol{C}\dot{\boldsymbol{D}}} &= \boldsymbol{\sigma}^\nu_{\phantom{\nu}\boldsymbol{C}\dot{\boldsymbol{D}}}\,X^\mu_{\phantom{\mu}\nu}\,\boldsymbol{\sigma}_\mu^{\phantom{\mu}\boldsymbol{A}\dot{\boldsymbol{B}}}.
\end{align}

Using the spin frames, we can write:
\begin{equation}
\boldsymbol{\sigma}^\mu_{\phantom{\mu}\boldsymbol{a}\dot{\boldsymbol{b}}} = \boldsymbol{\sigma}^\mu_{\phantom{\mu}\boldsymbol{A}\dot{\boldsymbol{B}}}\boldsymbol{\zeta}_{\boldsymbol{a}}^{\phantom{\boldsymbol{a}}\boldsymbol{A}}\overline{\boldsymbol{\zeta}}_{\boldsymbol{b}}^{\phantom{\boldsymbol{b}}\dot{\boldsymbol{B}}}.
\end{equation}

Finally, we want to define the covariant derivative. Since we have (see \cite{P1}) $g_{\boldsymbol{A}\dot{\boldsymbol{B}}\boldsymbol{C}\dot{\boldsymbol{D}}}= \boldsymbol{\epsilon}_{\boldsymbol{A}\boldsymbol{C}}\boldsymbol{\epsilon}_{\dot{\boldsymbol{B}}\dot{\boldsymbol{D}}}$, we have:
\begin{equation}
\nabla_\mu g_{\nu\rho}=0 \Rightarrow \nabla_\mu\,\boldsymbol{\epsilon}_{\boldsymbol{A}\boldsymbol{C}}=0, \quad \text{and} \quad \nabla_\mu\,\boldsymbol{\epsilon}_{\dot{\boldsymbol{B}}\dot{\boldsymbol{D}}}=0.
\end{equation}

For a generic multispinor $\boldsymbol{\kappa}_{\boldsymbol{A}}$, the covariant derivative is given by:
\begin{equation}
\nabla_\mu\,\boldsymbol{\kappa}_{\boldsymbol{A}} = \boldsymbol{\kappa}_{\boldsymbol{A},\mu} - \boldsymbol{\kappa}_{\boldsymbol{C}}\,\Gamma^{\boldsymbol{C}}_{\phantom{\boldsymbol{C}}\boldsymbol{A}\mu},
\end{equation}
where
\begin{equation}
\boldsymbol{\zeta}_{\boldsymbol{a}\boldsymbol{A};\mu}=-\boldsymbol{\zeta}_{\boldsymbol{a}\boldsymbol{C}}\,\Gamma^{\boldsymbol{C}}_{\phantom{\boldsymbol{C}}\boldsymbol{A}\mu}=\Gamma_{\boldsymbol{a}\boldsymbol{A}\mu}.
\end{equation}

The spin coefficients are given by \cite{NP}:
\begin{equation}
\begin{split}
\Gamma_{\boldsymbol{a}\boldsymbol{b}\boldsymbol{c}\dot{\boldsymbol{d}}}&=\boldsymbol{\zeta}_{\boldsymbol{a}\boldsymbol{A};\mu}\boldsymbol{\zeta}_{\boldsymbol{b}}^{\phantom{\boldsymbol{b}}\boldsymbol{A}}\,\boldsymbol{\sigma}^\mu_{\phantom{\mu}\boldsymbol{c}\dot{\boldsymbol{d}}}=\\
&=\dfrac{1}{2}\boldsymbol{\epsilon}^{\dot{\boldsymbol{p}}\dot{\boldsymbol{q}}}\boldsymbol{\sigma}^\nu_{\phantom{\nu}\boldsymbol{a}\dot{\boldsymbol{q}}}\boldsymbol{\sigma}^\mu_{\phantom{\mu}\boldsymbol{c}\dot{\boldsymbol{d}}}\boldsymbol{\sigma}_{\nu\boldsymbol{b}\dot{\boldsymbol{p}}}
\end{split}
\end{equation}
In the above equation we have used a notation we shall employ everywhere in the paper: we use $\boldsymbol{\epsilon}^{{\boldsymbol{p}}{\boldsymbol{q}}}$ to indicate a product of $\boldsymbol{\epsilon}$ spinors, one coming from each copy of $SU(2)$.

\section{The curvature multispinors}
\label{sec:curv}

As in the 4d case, also in higher dimensional spacetimes we have a particular relation between antisymmetric second rank tensors and symmetric second rank multispinors \cite{bade,P1}. In fact we have:
\begin{equation}
F_{\mu\nu}=-F_{\nu\mu} \Rightarrow F_{\boldsymbol{A}\dot{\boldsymbol{B}}\boldsymbol{C}\dot{\boldsymbol{D}}} = \dfrac{1}{2}\, \left[ \phi_{\boldsymbol{A}\boldsymbol{C}}\, \boldsymbol{\epsilon}_{\dot{\boldsymbol{B}}\dot{\boldsymbol{D}}} + \boldsymbol{\epsilon}_{\boldsymbol{A}\boldsymbol{C}}\, \overline{\phi}_{\dot{\boldsymbol{B}}\dot{\boldsymbol{D}}} \right]
\end{equation}
The procedure outlined in \cite{bade} can be repeated for any pair of antisymmetric indices. In particular also for the Riemann tensor $R_{\mu\nu\rho\sigma}$ which is antisymmetric in the pairs $\mu\nu$ and $\rho\sigma$. Then as in \cite{P1,witten}, we find:
\begin{equation}
\begin{split}
R_{\mu\nu\rho\sigma} \Rightarrow R_{\boldsymbol{A}\dot{\boldsymbol{E}}\boldsymbol{B}\dot{\boldsymbol{F}}\boldsymbol{C}\dot{\boldsymbol{G}}\boldsymbol{D}\dot{\boldsymbol{H}}} = \dfrac{1}{2}\, &\Big[ \chi_{\boldsymbol{ABCD}}\, \boldsymbol{\epsilon}_{\dot{\boldsymbol{E}}\dot{\boldsymbol{F}}}\, \boldsymbol{\epsilon}_{\dot{\boldsymbol{G}}\dot{\boldsymbol{H}}} + \boldsymbol{\epsilon}_{\boldsymbol{CD}}\, \phi_{\boldsymbol{AB}\dot{\boldsymbol{G}}\dot{\boldsymbol{H}}} \, \boldsymbol{\epsilon}_{\dot{\boldsymbol{E}}\dot{\boldsymbol{F}}} +\\
&+ \boldsymbol{\epsilon}_{\boldsymbol{AB}}\, \overline{\phi}_{\dot{\boldsymbol{E}}\dot{\boldsymbol{F}}\boldsymbol{CD}} \, \boldsymbol{\epsilon}_{\dot{\boldsymbol{G}}\dot{\boldsymbol{H}}} + \boldsymbol{\epsilon}_{\boldsymbol{AB}} \, \boldsymbol{\epsilon}_{\boldsymbol{CD}} \, \overline{\chi}_{\dot{\boldsymbol{E}}\dot{\boldsymbol{F}}\dot{\boldsymbol{G}}\dot{\boldsymbol{H}}} \Big]
\end{split}
\end{equation}

Using our multi-index notation, the scalars $\chi_{\boldsymbol{ABCD}}$ and $\phi_{\boldsymbol{AB}\dot{\boldsymbol{E}}\dot{\boldsymbol{F}}}$ satisfy the usual symmetry relations \cite{P1}:
\begin{align}
\chi_{\boldsymbol{ABCD}}&=\chi_{\boldsymbol{BACD}}=\chi_{\boldsymbol{ABDC}}=\chi_{\boldsymbol{CDAB}}\\
\phi_{\boldsymbol{AB}\dot{\boldsymbol{E}}\dot{\boldsymbol{F}}}&=\phi_{\boldsymbol{BA}\dot{\boldsymbol{E}}\dot{\boldsymbol{F}}}=\phi_{\boldsymbol{AB}\dot{\boldsymbol{F}}\dot{\boldsymbol{E}}}=\overline{\phi}_{\boldsymbol{EF}\dot{\boldsymbol{A}}\dot{\boldsymbol{B}}}
\end{align}

Riemann tensor also satisfies the symmetry relation $R_{\mu[\nu\rho\sigma]}=0$. In 4d (see \cite{P1}), one introduces the right dual:
\begin{equation}\label{eq:RD}
S_{\mu\nu\rho\sigma}=\dfrac{1}{2}\, \sqrt{-g}\, R_{\mu\nu}^{\phantom{\mu\nu}\alpha\beta}\,\epsilon_{\alpha\beta\rho\sigma},
\end{equation}
then the above symmetry relation becomes simply $S_{\mu\nu\rho}^{\phantom{\mu\nu\rho}\nu}=0$. In higher dimensions, \eqref{eq:RD} must be modified; for example in 6d, we have:
\begin{equation}
S_{\mu\nu\rho\sigma\tau\xi}=\dfrac{1}{2}\, \sqrt{-g}\, R_{\mu\nu}^{\phantom{\mu\nu}\alpha\beta}\,\epsilon_{\alpha\beta\rho\sigma\tau\xi},
\end{equation}
since now the fully antisymmetric symbol has 6 indices. We still have that $R_{\mu[\nu\rho\sigma]}=0=S_{\mu\nu\rho\sigma\tau}^{\phantom{\mu\nu\rho\sigma\tau}\nu}$. The treatment is analogous in higher dimensional spacetimes.

We introduce the tensor $\mathcal{E}^{\alpha\beta}_{\phantom{\alpha\beta}\rho\sigma\tau\xi} =\sqrt{-g}\, \epsilon_{\mu\nu\rho\sigma\tau\xi}\, g^{\alpha\mu}g^{\nu\beta}$. Then, in terms of multispinors, we have:
\begin{equation}
\begin{split}
\mathcal{E}^{\boldsymbol{E}\dot{\boldsymbol{F}}\boldsymbol{G}\dot{\boldsymbol{H}}}_{\boldsymbol{I}\dot{\boldsymbol{L}}\boldsymbol{M}\dot{\boldsymbol{N}}\boldsymbol{O}\dot{\boldsymbol{P}}\boldsymbol{Q}\dot{\boldsymbol{R}}}=&i\, \Bigg[ \left( \delta^{\boldsymbol{E}}_{\boldsymbol{I}}\delta^{\boldsymbol{G}}_{\boldsymbol{M}}\delta^{\dot{\boldsymbol{F}}}_{\dot{\boldsymbol{L}}}\delta^{\dot{\boldsymbol{H}}}_{\dot{\boldsymbol{N}}} - \delta^{\boldsymbol{E}}_{\boldsymbol{M}}\delta^{\boldsymbol{G}}_{\boldsymbol{I}}\delta^{\dot{\boldsymbol{F}}}_{\dot{\boldsymbol{N}}}\delta^{\dot{\boldsymbol{H}}}_{\dot{\boldsymbol{L}}} \right)\,\boldsymbol{\epsilon}_{\boldsymbol{OQ}}\,\boldsymbol{\epsilon}_{\dot{\boldsymbol{P}}\dot{\boldsymbol{R}}} +\\
&- \left( \delta^{\boldsymbol{E}}_{\boldsymbol{M}}\delta^{\boldsymbol{G}}_{\boldsymbol{O}}\delta^{\dot{\boldsymbol{F}}}_{\dot{\boldsymbol{N}}}\delta^{\dot{\boldsymbol{H}}}_{\dot{\boldsymbol{P}}} - \delta^{\boldsymbol{E}}_{\boldsymbol{O}}\delta^{\boldsymbol{G}}_{\boldsymbol{M}}\delta^{\dot{\boldsymbol{F}}}_{\dot{\boldsymbol{P}}}\delta^{\dot{\boldsymbol{H}}}_{\dot{\boldsymbol{N}}} \right)\,\boldsymbol{\epsilon}_{\boldsymbol{IQ}}\,\boldsymbol{\epsilon}_{\dot{\boldsymbol{L}}\dot{\boldsymbol{R}}}  +\\
&+ \left( \delta^{\boldsymbol{E}}_{\boldsymbol{O}}\delta^{\boldsymbol{G}}_{\boldsymbol{Q}}\delta^{\dot{\boldsymbol{F}}}_{\dot{\boldsymbol{P}}}\delta^{\dot{\boldsymbol{H}}}_{\dot{\boldsymbol{R}}} - \delta^{\boldsymbol{E}}_{\boldsymbol{Q}}\delta^{\boldsymbol{G}}_{\boldsymbol{O}}\delta^{\dot{\boldsymbol{F}}}_{\dot{\boldsymbol{R}}}\delta^{\dot{\boldsymbol{H}}}_{\dot{\boldsymbol{R}}} \right)\,\boldsymbol{\epsilon}_{\boldsymbol{IM}}\,\boldsymbol{\epsilon}_{\dot{\boldsymbol{L}}\dot{\boldsymbol{N}}}  \Bigg]
\end{split}
\end{equation}

From the above expression of the Riemann tensor in spinor form, we find the complicated formula:
\begin{equation}\label{eq:defS}
\begin{split}
S_{\boldsymbol{A}\dot{\boldsymbol{B}}\boldsymbol{C}\dot{\boldsymbol{D}}\boldsymbol{I}\dot{\boldsymbol{L}}\boldsymbol{M}\dot{\boldsymbol{N}}\boldsymbol{O}\dot{\boldsymbol{P}}\boldsymbol{Q}\dot{\boldsymbol{R}}}=\dfrac{i}{2} \, &\Bigg[ \Big( -\chi_{\boldsymbol{A}\boldsymbol{C}\boldsymbol{I}\boldsymbol{M}}\,\epsilon_{\dot{\boldsymbol{B}}\dot{\boldsymbol{D}}}\,\epsilon_{\dot{\boldsymbol{L}}\dot{\boldsymbol{N}}} + \epsilon_{\boldsymbol{I}\boldsymbol{M}}\, \phi_{\boldsymbol{A}\boldsymbol{C}\dot{\boldsymbol{L}}\dot{\boldsymbol{N}}}\,\epsilon_{\dot{\boldsymbol{B}}\dot{\boldsymbol{D}}} +\\
&- \epsilon_{\boldsymbol{A}\boldsymbol{C}}\, \overline{\phi}_{\dot{\boldsymbol{B}}\dot{\boldsymbol{D}}\boldsymbol{I}\boldsymbol{M}}\,\epsilon_{\dot{\boldsymbol{L}}\dot{\boldsymbol{N}}} + \overline{\chi}_{\dot{\boldsymbol{B}}\dot{\boldsymbol{D}}\dot{\boldsymbol{L}}\dot{\boldsymbol{N}}}\,\epsilon_{\boldsymbol{A}\boldsymbol{C}}\,\epsilon_{\boldsymbol{I}\boldsymbol{M}} \Big)\,\epsilon_{\boldsymbol{O}\boldsymbol{Q}} \,\epsilon_{\dot{\boldsymbol{P}}\dot{\boldsymbol{R}}} +\\
&-\Big( -\chi_{\boldsymbol{A}\boldsymbol{C}\boldsymbol{M}\boldsymbol{O}}\,\epsilon_{\dot{\boldsymbol{B}}\dot{\boldsymbol{D}}}\,\epsilon_{\dot{\boldsymbol{N}}\dot{\boldsymbol{P}}} + \epsilon_{\boldsymbol{M}\boldsymbol{O}}\, \phi_{\boldsymbol{A}\boldsymbol{C}\dot{\boldsymbol{N}}\dot{\boldsymbol{P}}}\,\epsilon_{\dot{\boldsymbol{B}}\dot{\boldsymbol{D}}} +\\
&- \epsilon_{\boldsymbol{A}\boldsymbol{C}}\, \overline{\phi}_{\dot{\boldsymbol{B}}\dot{\boldsymbol{D}}\boldsymbol{M}\boldsymbol{O}}\,\epsilon_{\dot{\boldsymbol{N}}\dot{\boldsymbol{P}}} + \overline{\chi}_{\dot{\boldsymbol{B}}\dot{\boldsymbol{D}}\dot{\boldsymbol{N}}\dot{\boldsymbol{P}}}\,\epsilon_{\boldsymbol{A}\boldsymbol{C}}\,\epsilon_{\boldsymbol{M}\boldsymbol{O}} \Big)\,\epsilon_{\boldsymbol{I}\boldsymbol{Q}} \,\epsilon_{\dot{\boldsymbol{L}}\dot{\boldsymbol{R}}} +\\ 
&+\Big( -\chi_{\boldsymbol{A}\boldsymbol{C}\boldsymbol{O}\boldsymbol{Q}}\,\epsilon_{\dot{\boldsymbol{B}}\dot{\boldsymbol{D}}}\,\epsilon_{\dot{\boldsymbol{P}}\dot{\boldsymbol{R}}} + \epsilon_{\boldsymbol{O}\boldsymbol{Q}}\, \phi_{\boldsymbol{A}\boldsymbol{C}\dot{\boldsymbol{P}}\dot{\boldsymbol{R}}}\,\epsilon_{\dot{\boldsymbol{B}}\dot{\boldsymbol{D}}} +\\
&- \epsilon_{\boldsymbol{A}\boldsymbol{C}}\, \overline{\phi}_{\dot{\boldsymbol{B}}\dot{\boldsymbol{D}}\boldsymbol{O}\boldsymbol{Q}}\,\epsilon_{\dot{\boldsymbol{P}}\dot{\boldsymbol{R}}} + \overline{\chi}_{\dot{\boldsymbol{B}}\dot{\boldsymbol{D}}\dot{\boldsymbol{P}}\dot{\boldsymbol{R}}}\,\epsilon_{\boldsymbol{A}\boldsymbol{C}}\,\epsilon_{\boldsymbol{O}\boldsymbol{Q}} \Big)\,\epsilon_{\boldsymbol{I}\boldsymbol{M}} \,\epsilon_{\dot{\boldsymbol{L}}\dot{\boldsymbol{N}}}\Bigg].
\end{split}
\end{equation}

Now the expression $S_{\mu\nu\rho\sigma\tau}^{\phantom{\mu\nu\rho\sigma\tau}\nu}=0$ reduces to:
\begin{equation}
\begin{split}
&\Big( -\chi_{\boldsymbol{A}\boldsymbol{C}\boldsymbol{I}\boldsymbol{M}}\,\epsilon_{\dot{\boldsymbol{B}}\dot{\boldsymbol{D}}}\,\epsilon_{\dot{\boldsymbol{L}}\dot{\boldsymbol{N}}} + \epsilon_{\boldsymbol{I}\boldsymbol{M}}\, \phi_{\boldsymbol{A}\boldsymbol{C}\dot{\boldsymbol{L}}\dot{\boldsymbol{N}}}\,\epsilon_{\dot{\boldsymbol{B}}\dot{\boldsymbol{D}}} +\\
-& \epsilon_{\boldsymbol{A}\boldsymbol{C}}\, \overline{\phi}_{\dot{\boldsymbol{B}}\dot{\boldsymbol{D}}\boldsymbol{I}\boldsymbol{M}}\,\epsilon_{\dot{\boldsymbol{L}}\dot{\boldsymbol{N}}} + \overline{\chi}_{\dot{\boldsymbol{B}}\dot{\boldsymbol{D}}\dot{\boldsymbol{L}}\dot{\boldsymbol{N}}}\,\epsilon_{\boldsymbol{A}\boldsymbol{C}}\,\epsilon_{\boldsymbol{I}\boldsymbol{M}} \Big)\,\delta_{\boldsymbol{O}}^{\boldsymbol{C}} \,\delta_{\dot{\boldsymbol{P}}}^{\dot{\boldsymbol{D}}} +\\
-&\Big( -\chi_{\boldsymbol{A}\boldsymbol{C}\boldsymbol{M}\boldsymbol{O}}\,\epsilon_{\dot{\boldsymbol{B}}\dot{\boldsymbol{D}}}\,\epsilon_{\dot{\boldsymbol{N}}\dot{\boldsymbol{P}}} + \epsilon_{\boldsymbol{M}\boldsymbol{O}}\, \phi_{\boldsymbol{A}\boldsymbol{C}\dot{\boldsymbol{N}}\dot{\boldsymbol{P}}}\,\epsilon_{\dot{\boldsymbol{B}}\dot{\boldsymbol{D}}} +\\
-& \epsilon_{\boldsymbol{A}\boldsymbol{C}}\, \overline{\phi}_{\dot{\boldsymbol{B}}\dot{\boldsymbol{D}}\boldsymbol{M}\boldsymbol{O}}\,\epsilon_{\dot{\boldsymbol{N}}\dot{\boldsymbol{P}}} + \overline{\chi}_{\dot{\boldsymbol{B}}\dot{\boldsymbol{D}}\dot{\boldsymbol{N}}\dot{\boldsymbol{P}}}\,\epsilon_{\boldsymbol{A}\boldsymbol{C}}\,\epsilon_{\boldsymbol{M}\boldsymbol{O}} \Big)\,\delta_{\boldsymbol{I}}^{\boldsymbol{C}} \,\delta_{\dot{\boldsymbol{L}}}^{\dot{\boldsymbol{D}}} +\\ 
+&\Big( -\chi_{\boldsymbol{A}\boldsymbol{C}\boldsymbol{O}}^{\phantom{\boldsymbol{A}\boldsymbol{C}\boldsymbol{O}}\boldsymbol{C}}\,\epsilon_{\dot{\boldsymbol{B}}\dot{\boldsymbol{D}}}\,\delta_{\dot{\boldsymbol{P}}}^{\dot{\boldsymbol{D}}} + \delta_{\boldsymbol{O}}^{\boldsymbol{C}}\, \phi_{\boldsymbol{A}\boldsymbol{C}\dot{\boldsymbol{P}}}^{\phantom{\boldsymbol{A}\boldsymbol{C}\dot{\boldsymbol{P}}}\dot{\boldsymbol{D}}}\,\epsilon_{\dot{\boldsymbol{B}}\dot{\boldsymbol{D}}} +\\
-& \epsilon_{\boldsymbol{A}\boldsymbol{C}}\, \overline{\phi}_{\dot{\boldsymbol{B}}\dot{\boldsymbol{D}}\boldsymbol{O}}^{\phantom{\dot{\boldsymbol{B}}\dot{\boldsymbol{D}}\boldsymbol{O}}\boldsymbol{C}}\,\delta_{\dot{\boldsymbol{P}}}^{\dot{\boldsymbol{D}}} + \overline{\chi}_{\dot{\boldsymbol{B}}\dot{\boldsymbol{D}}\dot{\boldsymbol{P}}}^{\phantom{\dot{\boldsymbol{B}}\dot{\boldsymbol{D}}\dot{\boldsymbol{P}}}\dot{\boldsymbol{D}}}\,\epsilon_{\boldsymbol{A}\boldsymbol{C}}\,\delta_{\boldsymbol{O}}^{\boldsymbol{C}} \Big)\,\epsilon_{\boldsymbol{I}\boldsymbol{M}} \,\epsilon_{\dot{\boldsymbol{L}}\dot{\boldsymbol{N}}} = 0
\end{split}
\end{equation}

The expression in the parenthesis must be separately equated to zero. The last parenthesis gives (compare with \cite{P1}):
\begin{equation}
\chi_{\boldsymbol{ABC}}^{\phantom{\boldsymbol{ABC}}\boldsymbol{B}}= \Lambda\, \boldsymbol{\epsilon}_{\boldsymbol{AC}}, \qquad \Lambda= \dfrac{1}{2}\, \chi_{\boldsymbol{AB}}^{\phantom{\boldsymbol{AB}}\boldsymbol{AB}} = \dfrac{1}{2}\, \overline{\chi}_{\dot{\boldsymbol{A}}\dot{\boldsymbol{B}}}^{\phantom{\boldsymbol{AB}}\dot{\boldsymbol{A}}\dot{\boldsymbol{B}}}
\end{equation}
where $\Lambda$ is a real scalar. The other lines give additional relations that must be satisfied by the curvature spinors in higher dimensions. Separating the parts symmetric in $\dot{\boldsymbol{L}}\dot{\boldsymbol{N}}$ or in $\dot{\boldsymbol{N}}\dot{\boldsymbol{P}}$ and $\boldsymbol{AC}$ from that antisymmetric, we find the following relations:
\begin{equation}
\chi_{\boldsymbol{A}\boldsymbol{O}\boldsymbol{I}\boldsymbol{M}}=0, \quad \phi_{\boldsymbol{A}\boldsymbol{O}\dot{\boldsymbol{L}}\dot{\boldsymbol{N}}}=0.
\end{equation}

Finally, we derive the Ricci tensor $R_{\mu\nu}=R^\sigma_{\phantom{\sigma}\mu\sigma\nu}$; in spinor form it is given by (compare to \cite{P1}):
\begin{equation}
\begin{split}
R_{\boldsymbol{A}\dot{\boldsymbol{B}}\boldsymbol{C}\dot{\boldsymbol{D}}} &= \dfrac{1}{2} \, \Big[ \chi_{\boldsymbol{EA}\phantom{\boldsymbol{E}}\boldsymbol{C}}^{\phantom{\boldsymbol{EA}}\boldsymbol{E}}\, \boldsymbol{\epsilon}_{\dot{\boldsymbol{B}}\dot{\boldsymbol{D}}} - 2\phi_{\boldsymbol{AC}\dot{\boldsymbol{B}}\dot{\boldsymbol{D}}} + \boldsymbol{\epsilon}_{\boldsymbol{AC}}\,\overline{\chi}_{\dot{\boldsymbol{F}}\dot{\boldsymbol{B}}\phantom{\boldsymbol{F}}\dot{\boldsymbol{D}}}^{\phantom{\boldsymbol{FB}}\dot{\boldsymbol{F}}} \Big]=\\
&=\Lambda\, \boldsymbol{\epsilon}_{\boldsymbol{AC}}\,\boldsymbol{\epsilon}_{\dot{\boldsymbol{B}}\dot{\boldsymbol{D}}} - \phi_{\boldsymbol{AC}\dot{\boldsymbol{B}}\dot{\boldsymbol{D}}}.
\end{split}
\end{equation}
The Ricci scalar is given by $R=4\,\Lambda$.

The above relations mean that, in vacuum, we have:
\begin{equation}
\Lambda=0, \qquad \phi_{\boldsymbol{AB}\dot{\boldsymbol{C}}\dot{\boldsymbol{D}}}=0,
\end{equation}
then $\chi_{\boldsymbol{ABC}}^{\phantom{\boldsymbol{ABC}}\boldsymbol{B}}=0$, so that $\chi$ is completely symmetric in all of its indices in vacuum. Then, following \cite{P1}, we can introduce the Weyl scalar $\psi_{\boldsymbol{ABCD}}$ as follows (compare to \cite{RP}):
\begin{equation}
\chi_{\boldsymbol{ABCD}} = \psi_{\boldsymbol{ABCD}} + \dfrac{\Lambda}{3}\, \Big[ \boldsymbol{\epsilon}_{\boldsymbol{AC}}\boldsymbol{\epsilon}_{\boldsymbol{BD}} + \boldsymbol{\epsilon}_{\boldsymbol{AD}}\boldsymbol{\epsilon}_{\boldsymbol{BC}} \Big].
\end{equation}

\section{The Bianchi identities}
\label{sec:bia}

The Riemann tensor also satisfies the Bianchi identities $\nabla_{[\tau}R_{\mu\nu]\rho\sigma}=0$. In 6d, in terms of the tensor $S_{\mu\nu\rho\sigma\tau\xi}$ defined above, they can be written as $\nabla^\xi S_{\mu\nu\rho\sigma\tau\xi}=0$. Using spinor notation and the expression \eqref{eq:defS}, we can derive the 6d Bianchi identities in spinor form:
\begin{equation}\label{eq:Bianchi}
\begin{split}
&\nabla_{\boldsymbol{O}\dot{\boldsymbol{P}}}\,\Big( -\chi_{\boldsymbol{A}\boldsymbol{C}\boldsymbol{I}\boldsymbol{M}} \,\epsilon_{\dot{\boldsymbol{L}}\dot{\boldsymbol{N}}} + \epsilon_{\boldsymbol{I}\boldsymbol{M}}\, \phi_{\boldsymbol{A}\boldsymbol{C}\dot{\boldsymbol{L}}\dot{\boldsymbol{N}}} \Big) - \nabla_{\boldsymbol{I}\dot{\boldsymbol{L}}}\,\Big( -\chi_{\boldsymbol{A}\boldsymbol{C}\boldsymbol{M}\boldsymbol{O}}\,\epsilon_{\dot{\boldsymbol{N}}\dot{\boldsymbol{P}}} + \epsilon_{\boldsymbol{M}\boldsymbol{O}}\, \phi_{\boldsymbol{A}\boldsymbol{C}\dot{\boldsymbol{N}}\dot{\boldsymbol{P}}} \Big)+\\
-&\, \Big( \nabla^{\boldsymbol{Q}}_{\phantom{\boldsymbol{Q}}\dot{\boldsymbol{P}}}\,\chi_{\boldsymbol{A}\boldsymbol{C}\boldsymbol{O}\boldsymbol{Q}} - \nabla_{\boldsymbol{O}}^{\phantom{\boldsymbol{O}}\dot{\boldsymbol{R}}}\,\phi_{\boldsymbol{A}\boldsymbol{C}\dot{\boldsymbol{P}}\dot{\boldsymbol{R}}} \Big) = 0.
\end{split}
\end{equation}

The last line gives the usual Bianchi identities (see \cite{P1}):
\begin{equation}
\nabla^{\boldsymbol{Q}}_{\phantom{\boldsymbol{Q}}\dot{\boldsymbol{P}}}\,\chi_{\boldsymbol{A}\boldsymbol{C}\boldsymbol{O}\boldsymbol{Q}} = \nabla_{(\boldsymbol{O}}^{\phantom{\boldsymbol{O}}\dot{\boldsymbol{R}}}\,\phi_{\boldsymbol{A}\boldsymbol{C})\dot{\boldsymbol{P}}\dot{\boldsymbol{R}}}.
\end{equation}
Using the definition of $\psi_{\boldsymbol{ABCD}}$ and $\Lambda$ given above, this expression can be written as two separate relations (see also \cite{NP}):
\begin{align}
\nabla^{\boldsymbol{D}}_{\phantom{\boldsymbol{D}}\dot{\boldsymbol{G}}}\psi_{\boldsymbol{ABCD}}&=\nabla_{(\boldsymbol{C}}^{\phantom{\boldsymbol{C}}\dot{\boldsymbol{H}}}\phi_{\boldsymbol{AB})\dot{\boldsymbol{G}}\dot{\boldsymbol{H}}},\\
\nabla^{\boldsymbol{A}\dot{\boldsymbol{G}}}\phi_{\boldsymbol{AB}\dot{\boldsymbol{G}}\dot{\boldsymbol{H}}} &= -3\, \nabla_{\boldsymbol{B}\dot{\boldsymbol{H}}}\Lambda.
\end{align}

The first line of \eqref{eq:Bianchi} gives additional Bianchi identities valid in higher dimension. Separating the part symmetric in $\dot{\boldsymbol{L}}\dot{\boldsymbol{N}}$ or $\dot{\boldsymbol{N}}\dot{\boldsymbol{P}}$ from that antisymmetric, we find:
\begin{align}
\nabla_{\boldsymbol{O}\dot{\boldsymbol{P}}}\,\chi_{\boldsymbol{A}\boldsymbol{C}\boldsymbol{I}\boldsymbol{M}} &=0,\\
\nabla_{\boldsymbol{O}\dot{\boldsymbol{P}}}\,\phi_{\boldsymbol{A}\boldsymbol{C}\dot{\boldsymbol{L}}\dot{\boldsymbol{N}}} &=0,\\
\nabla_{\boldsymbol{I}\dot{\boldsymbol{L}}}\,\chi_{\boldsymbol{A}\boldsymbol{C}\boldsymbol{M}\boldsymbol{O}} &=0,\\
\nabla_{\boldsymbol{I}\dot{\boldsymbol{L}}}\,\phi_{\boldsymbol{A}\boldsymbol{C}\dot{\boldsymbol{N}}\dot{\boldsymbol{P}}} &=0.
\end{align}

\section{The Newman-Penrose formalism}
\label{sec:NP}

In this Section,  we follow closely \cite{NP} and rederive some of their results in our formalism, in particular, we introduce a null vielbein, we calculate the spin coefficients, the Weyl and Ricci scalars and the commutation relations between the derivatives. We also calculate the sixty derivatives of the spin coefficients, but their expression is long, so we report them in the Appendix \ref{app:der}.
\subsection{Null vielbein}

We can associate a null vielbein to our spin frame. In the 6d case we have:
\begin{align}\label{eq:rel1}
l^\mu&=[\boldsymbol{\omicron}_1^{A_1},\boldsymbol{\omicron}_2^{B_2}][\boldsymbol{\omicron}_1^{\dot{A}_1},\boldsymbol{\omicron}_2^{\dot{B}_2}] = \boldsymbol{\sigma}^\mu_{\phantom{\mu}0_10_2\dot{0}_1\dot{0}_2}\\
n^\mu&=[\boldsymbol{\iota}_1^{A_1},\boldsymbol{\iota}_2^{B_2}][\boldsymbol{\iota}_1^{\dot{A}_1},\boldsymbol{\iota}_2^{\dot{B}_2}] = \boldsymbol{\sigma}^\mu_{\phantom{\mu}1_11_2\dot{1}_1\dot{1}_2}\\
m_1^\mu&=[\boldsymbol{\omicron}_1^{A_1},\boldsymbol{\iota}_2^{B_2}][\boldsymbol{\omicron}_1^{\dot{A}_1},\boldsymbol{\omicron}_2^{\dot{B}_2}] = \boldsymbol{\sigma}^\mu_{\phantom{\mu}0_11_2\dot{0}_1\dot{0}_2}\\
m_2^\mu&=[\boldsymbol{\iota}_1^{A_1},\boldsymbol{\omicron}_2^{B_2}][\boldsymbol{\iota}_1^{\dot{A}_1},\boldsymbol{\iota}_2^{\dot{B}_2}] = \boldsymbol{\sigma}^\mu_{\phantom{\mu}1_10_2\dot{1}_1\dot{1}_2}\\
m_3^\mu&=[\boldsymbol{\omicron}_1^{A_1},\boldsymbol{\omicron}_2^{B_2}][\boldsymbol{\iota}_1^{\dot{A}_1},\boldsymbol{\omicron}_2^{\dot{B}_2}] = \boldsymbol{\sigma}^\mu_{\phantom{\mu}0_10_2\dot{1}_1\dot{0}_2}\\\label{eq:rel_fin}
m_4^\mu&=[\boldsymbol{\iota}_1^{A_1},\boldsymbol{\iota}_2^{B_2}][\boldsymbol{\omicron}_1^{\dot{A}_1},\boldsymbol{\iota}_2^{\dot{B}_2}] = \boldsymbol{\sigma}^\mu_{\phantom{\mu}1_11_2\dot{0}_1\dot{1}_2}
\end{align}

We have the relations:
\begin{equation}
l^\mu l_\mu=n^\mu n_\mu=m_i^\mu m_{i\mu}=0 \qquad i=\{1,2,3,4\}
\end{equation}
indicating that indeed our vielbein is null and, moreover, we have:
\begin{equation}
1=l^\mu n_\mu = -m_1^\mu m_{2\mu} = -m_3^\mu m_{4\mu}
\end{equation}
all other contractions giving 0.

The extension to higher dimensional spacetime is trivial. For the odd dimensional case, we notice that it is not possible to have a full null vielbein, the spare vector must be space-like with unit norm. This indicates that our formalism only works for even-dimensional spacetimes.

We introduce the notation
\begin{equation}
z_A^\mu = \{l^\mu,n^\mu,m_1^\mu,m_2^\mu,m_3^\mu,m_4^\mu\}, \qquad A=\{1,2,3,4,5,6\}.
\end{equation}

The spin coefficients are given by \cite{NP}:
\begin{equation}\label{eq:SCT}
\gamma^{\phantom{A}BC}_A=z_{A\mu;\nu} z^{B\mu}z^{C\nu}
\end{equation}
with the symmetry
\begin{equation}
\gamma^{ABC}=-\gamma^{BAC}
\end{equation}

Finally, we introduce the derivatives:
\begin{align}
D\varphi&=\partial_{0_10_2\dot{0}_1\dot{0}_2}\varphi=l^\mu\nabla_\mu\varphi\\ \Delta\varphi&=\partial_{1_11_2\dot{1}_1\dot{1}_2}\varphi=n^\mu\nabla_\mu\varphi\\
\delta_1\varphi&=\partial_{0_11_2\dot{0}_1\dot{0}_2}\varphi=m_3^\mu\nabla_\mu\varphi\\
\delta_2\varphi&=\partial_{1_10_2\dot{1}_1\dot{1}_2}\varphi=m_4^\mu\nabla_\mu\varphi\\ \delta_3\varphi&=\partial_{0_10_2\dot{1}_1\dot{0}_2}\varphi=m_1^\mu\nabla_\mu\varphi\\
\delta_4\varphi&=\partial_{1_11_2\dot{0}_1\dot{1}_2}\varphi=m_2^\mu\nabla_\mu\varphi
\end{align}

\subsection{Null rotation}
The null rotations in $d>4$ dimensions are similar to those of 4d \cite{JN}. In the 6d case we are considering, we have:
\begin{align}\label{eq:tipo1}
\tilde{l}^\mu   &=l^\mu,\\
\tilde{m}_i^\mu &=m_i^\mu+a\,l^\mu,\\
\tilde{n}^\mu   &=n^\mu+a m_2^\mu+\overline{a}m_1^\mu+a m_4^\mu+\overline{a}m_3^\mu+a\overline{a}\,l^\mu
\end{align}
or
\begin{align}\label{eq:tipo2}
\tilde{n}^\mu   &=n^\mu,\\
\tilde{m}_i^\mu &=m_i^\mu+b\,n^\mu,\\
\tilde{l}^\mu   &=l^\mu+b m_2^\mu+\overline{b}m_1^\mu+b m_4^\mu+\overline{b}m_3^\mu+b\overline{b}\,l^\mu
\end{align}
or
\begin{align}
\tilde{l}^\mu   &=\lambda\,l^\mu,\\
\tilde{n}_i^\mu &=\lambda^{-1}n^\mu,\\
\tilde{m_i}^\mu &=\exp(i\alpha)m_i^\mu, \quad i=\{1,2\},\\
\tilde{m_j}^\mu &=\exp(i\beta)m_j^\mu, \quad j=\{3,4\},
\end{align}

\subsection{Spin coefficients}
Following \cite{NP}, we introduce the spin coefficients, focusing on 6d spacetimes according to the formula:
\begin{equation}
\Gamma_{a_1a_2b_1b_2c_1c_2\dot{d}_1\dot{d}_2}=\dfrac{1}{2} \boldsymbol{\epsilon}^{\dot{p}_1\dot{q}_1} \boldsymbol{\epsilon}^{\dot{p}_2\dot{q}_2}\boldsymbol{\sigma}^\mu_{\phantom{\mu}a_1a_2\dot{q}_1\dot{q}_2}\boldsymbol{\sigma}^\nu_{\phantom{\mu}c_1c_2\dot{d}_1\dot{d}_2}\boldsymbol{\sigma}_{\mu b_1b_2\dot{p}_1\dot{p}_2;\nu}.
\end{equation}
There are a total of 24 possible symbols, as can be deduced from table \ref{tab:spin}.

\begin{table}[h!]
\centering
\begin{tabular}{c||c|c|c|c|c|c}
&\multicolumn{4}{c}{$a_1a_2b_1b_2$}\\
\cline{2-5}
$c_1c_2\dot{d}_1\dot{d}_2$&$1_10_20_10_2$, or& $0_11_21_10_2$, or,& $0_10_21_11_2$, or& $0_11_21_11_2$, or \\
&or $0_10_21_10_2$ &or $1_10_20_11_2$& or $1_11_20_10_2$& or $1_11_20_11_2$ \\
\hline
$0_10_2\dot{0}_1\dot{0}_2$&$\alpha_1$&$\alpha_2$&$\alpha_3$&$\alpha_4$\\
$0_10_2\dot{1}_1\dot{0}_2$&$\beta_1$&$\beta_2$&$\beta_3$&$\beta_4$\\%
$1_11_2\dot{0}_1\dot{1}_2$&$\gamma_1$&$\gamma_2$&$\gamma_3$&$\gamma_4$\\
$0_11_2\dot{0}_1\dot{0}_2$&$\xi_1$&$\xi_2$&$\xi_3$&$\xi_4$\\
$1_10_2\dot{1}_1\dot{1}_2$&$\epsilon_1$&$\epsilon_2$&$\epsilon_3$&$\epsilon_4$\\
$1_11_2\dot{1}_1\dot{1}_2$&$\zeta_1$&$\zeta_2$&$\zeta_3$&$\zeta_4$\\
\end{tabular}
\caption{The spin coefficients indices and their definition.}\label{tab:spin}
\end{table}

We can establish the following relations between the above spin coefficients and the one given by \eqref{eq:SCT}:
\begin{align}
\Gamma_{1_10_20_10_20_10_2\dot{0}_1\dot{0}_2}&=\alpha_1=\gamma_{141}=l_{\mu;\nu}m_2^\mu l^\nu,\\
\Gamma_{0_11_21_10_20_10_2\dot{0}_1\dot{0}_2}&=\alpha_2=\gamma_{431}=m_{2\mu;\nu}m_1^\mu l^\nu,\\
\Gamma_{0_10_21_11_20_10_2\dot{0}_1\dot{0}_2}&=\alpha_3=\dfrac{1}{2}(\gamma_{651}+\gamma_{211})=\dfrac{1}{2}(m_{4\mu;\nu}m_3^\mu l^\nu+n_{\mu;\nu}l^\mu l^\nu),\\
\Gamma_{0_11_21_11_20_10_2\dot{0}_1\dot{0}_2}&=\alpha_4=\gamma_{321}=m_{1\mu;\nu}n^\mu l^\nu,\\
\Gamma_{1_10_20_10_20_10_2\dot{1}_1\dot{0}_2}&=\beta_1=\gamma_{145}=l_{\mu;\nu}m_2^\mu m_3^\nu,\\
\Gamma_{0_11_11_10_20_10_2\dot{1}_1\dot{0}_2}&=\beta_2=\gamma_{435}=m_{2\mu;\nu}m_1^\mu m_3^\nu,\\
\Gamma_{0_10_21_11_20_10_2\dot{1}_1\dot{0}_2}&=\beta_3=\dfrac{1}{2}(\gamma_{655}+\gamma_{215})=\dfrac{1}{2}(m_{4\mu;\nu}m_3^\mu m_3^\nu+n_{\mu;\nu}l^\mu m_3^\nu),\\
\Gamma_{0_11_21_11_20_10_2\dot{1}_1\dot{0}_2}&=\beta_4=\gamma_{325}=m_{1\mu;\nu}n^\mu m_3^\nu,\\
\Gamma_{1_10_20_10_21_11_2\dot{0}_1\dot{1}_2}&=\gamma_1=\gamma_{146}=l_{\mu;\nu}m_2^\mu m_4^\nu,\\
\Gamma_{0_11_21_10_21_11_2\dot{0}_1\dot{1}_2}&=\gamma_2=\gamma_{436}=m_{2\mu;\nu}m_1^\mu m_4^\nu,\\
\Gamma_{0_10_21_11_21_11_2\dot{0}_1\dot{1}_2}&=\gamma_3=\dfrac{1}{2}(\gamma_{656}+\gamma_{216})=\dfrac{1}{2}(m_{4\mu;\nu}m_3^\mu m_4^\nu+n_{\mu;\nu}l^\mu m_4^\nu),\\
\Gamma_{0_11_21_11_21_11_2\dot{0}_1\dot{1}_2}&=\gamma_4=\gamma_{326}=m_{1\mu;\nu}n^\mu m_4^\nu,\\
\Gamma_{1_10_20_10_20_10_2\dot{1}_1\dot{0}_2}&=\xi_1=\gamma_{143}=l_{\mu;\nu}m_2^\mu m_1^\nu,\\
\Gamma_{0_11_21_10_20_10_2\dot{1}_1\dot{0}_2}&=\xi_2=\gamma_{433}=m_{2\mu;\nu}m_1^\mu m_1^\nu,\\
\Gamma_{0_10_21_11_20_10_2\dot{1}_1\dot{0}_2}&=\xi_3=\dfrac{1}{2}(\gamma_{653}+\gamma_{213})=\dfrac{1}{2}(m_{4\mu;\nu}m_3^\mu m_1^\nu+n_{\mu;\nu}l^\mu m_1^\nu),\\
\Gamma_{0_11_21_11_20_10_2\dot{1}_1\dot{0}_2}&=\xi_4=\gamma_{323}=m_{1\mu;\nu}n^\mu m_1^\nu,\\
\Gamma_{1_10_20_10_21_11_2\dot{0}_1\dot{1}_2}&=\epsilon_1=\gamma_{144}=l_{\mu;\nu}m_2^\mu m_2^\nu,\\
\Gamma_{0_11_21_10_21_11_2\dot{0}_1\dot{1}_2}&=\epsilon_2=\gamma_{434}=m_{2\mu;\nu}m_1^\mu m_2^\nu,\\
\Gamma_{0_10_21_11_21_11_2\dot{0}_1\dot{1}_2}&=\epsilon_3=\dfrac{1}{2}(\gamma_{654}+\gamma_{214})=\dfrac{1}{2}(m_{4\mu;\nu}m_3^\mu m_2^\nu+n_{\mu;\nu}l^\mu m_2^\nu),\\
\Gamma_{0_11_21_11_21_11_2\dot{0}_1\dot{1}_2}&=\epsilon_4=\gamma_{324}=m_{1\mu;\nu}n^\mu m_2^\nu,\\
\Gamma_{1_10_20_10_21_11_2\dot{1}_1\dot{1}_2}&=\zeta_1=\gamma_{142}=l_{\mu;\nu}m_2^\mu n^\nu,\\
\Gamma_{0_11_21_10_21_11_2\dot{1}_1\dot{1}_2}&=\zeta_2=\gamma_{432}=m_{2\mu;\nu}m_1^\mu n^\nu,\\
\Gamma_{0_10_21_11_21_11_2\dot{1}_1\dot{1}_2}&=\zeta_3=\dfrac{1}{2}(\gamma_{652}+\gamma_{212})=\dfrac{1}{2}(m_{4\mu;\nu}m_3^\mu n^\nu+n_{\mu;\nu}l^\mu n^\nu),\\
\Gamma_{0_11_21_11_21_11_2\dot{1}_1\dot{1}_2}&=\zeta_4=\gamma_{322}=m_{1\mu;\nu}n^\mu n^\nu.
\end{align}

\subsection{Physical meaning of some of the spin coefficients}
We first consider
\begin{equation}
l_{\mu;\nu}\,l^\nu = -\alpha_1\,m_1^\mu - \overline{\alpha}_1 m_2^\mu +(\alpha_3+\overline{\alpha}_3)\,l^\mu
\end{equation}
So we find that $\alpha_3+\overline{\alpha}_3$ can be made to vanish with a rescaling of $l^\mu$, while if $\alpha_1=0$, $l^\mu$ is tangent to a geodesic. We might want our veilbein to be propagated parallel to the geodetic now defined, so have to impose:
\begin{equation}
\alpha_1=\alpha_3=0.
\end{equation}

In the case of a geodesic congruence, it can be seen, following for example \cite{poisson}, that the shear is given by:
\begin{equation}
\sigma=l_{\mu;\nu}m_2^\mu m_2^\nu=\epsilon_1,
\end{equation}
while the divergence (the sum of the expansion $\Theta$ and of the rotation $i\omega$) is:
\begin{equation}
\rho=l_{\mu;\nu}m_2^\mu m_1^\nu=\xi_1.
\end{equation}
In 6d spacetimes, there are two more shears:
\begin{equation}
\sigma_1=l_{\mu;\nu}m_2^\mu m_3^\nu=\beta_1, \qquad \sigma_2=l_{\mu;\nu}m_2^\mu m_4^\nu=\gamma_1.
\end{equation}

We next consider
\begin{equation}
l_{\mu;\nu}n^\nu= -\zeta_1\,m_1^\mu - \overline{\zeta}_1 m_2^\mu + (\zeta_3+\overline{\zeta}_3)\,l^\mu
\end{equation}
Again, with a rescaling of $l^\mu$ we can make $\zeta_3+\overline{\zeta}_3=0$. 

In the case of a hypersurface orthogonal geodesic congruence, i.e. a geodesics congruence proportional to a gradient field, we have that $\xi_1=\overline{\xi}_1$, i.e. it is real. If the congruence is equal to a gradient field, we have also $\zeta_1=\overline{\xi}_2+\epsilon_3$. 

For a $n^\mu$ congruence, instead of the $l^\mu$ congruence considered above, we have that the analogues of $\alpha_1$ is $\overline{\zeta}_4$ and the analogues of $\epsilon_1$ and $\xi_1$ are, respectively, $\overline{\xi}_4$ and $\epsilon_4$.

\subsection{Weyl scalars and Ricci tensor in vielbein form}
From the definition of the Weyl scalars, we have the following relations between multispinor and vielbein form:
\begin{align}
\psi_0&=-C_{1414}=-C_{\mu\nu\rho\sigma}l^\mu m_2^\nu l^\rho m_2^\sigma=\psi_{0_10_20_10_21_10_21_10_2},\\
\psi_1&=-C_{1443}=-C_{\mu\nu\rho\sigma}l^\mu m_2^\nu m_2^\rho m_1^\sigma=\psi_{0_10_21_10_21_10_20_11_2},\\
\psi_2&=-C_{1456}=-C_{\mu\nu\rho\sigma}l^\mu m_2^\nu m_3^\rho m_4^\sigma=\psi_{0_10_20_10_21_10_21_11_2},\\
\psi_3&=-C_{4343}=-C_{\mu\nu\rho\sigma}m_2^\mu m_1^\nu m_2^\rho m_3^\sigma=\psi_{1_10_20_11_21_10_20_11_2},\\
\psi_4&=-\dfrac{1}{2}(C_{6543}+C_{1234})=\\
&=-\dfrac{1}{2}C_{\mu\nu\rho\sigma}(m_2^\mu m_1^\nu m_4^\rho m_3^\sigma + l^\mu n^\nu m_1^\rho m_2^\sigma=\psi_{1_10_20_11_20_10_21_11_2},\\\nonumber
\psi_5&=-\dfrac{1}{3}(C_{6363}+C_{1256}+C_{1212})=\\
&=-\dfrac{1}{3}C_{\mu\nu\rho\sigma}(m_4^\mu m_1^\nu m_4^\rho m_1^\sigma+l^\mu n^\nu m_3^\rho m_4^\sigma+l^\mu n^\nu l^\rho n^\sigma)=\psi_{0_10_21_11_20_10_21_11_2},\\
\psi_6&=-C_{2323} = -C_{\mu\nu\rho\sigma}n^\mu m_1^\nu n^\rho m_1^\sigma = \psi_{1_11_20_11_21_11_20_11_2},\\
\psi_7&=-C_{2312}=-C_{\mu\nu\rho\sigma}\,n^\mu m_1^\nu l^\rho n^\sigma = \psi_{1_11_20_10_21_11_20_10_2},\\
\psi_8&=+C_{2343}=C_{\mu\nu\rho\sigma}\,n^\mu m_1^\nu m_2^\rho m_1^\sigma = \psi_{1_11_20_11_21_10_20_11_2}.
\end{align}
With the usual procedure, one can check that $\psi_6$ is linked to the gravitational waves amplitude.

The Ricci {tensors} are given by:
\begin{align}
\Phi_{0000}&=\phi_{0_10_20_10_2\dot{0}_1\dot{0}_2\dot{0}_1\dot{0}_2}=R_{11},\\
\Phi_{0100}&=\phi_{0_11_20_10_2\dot{0}_1\dot{0}_2\dot{0}_1\dot{0}_2}=R_{13},\\
\Phi_{0200}&=\phi_{0_11_20_11_2\dot{0}_1\dot{0}_2\dot{0}_1\dot{0}_2}=R_{33},\\
\Phi_{0010}&=\phi_{0_10_20_10_2\dot{1}_1\dot{0}_2\dot{0}_1\dot{0}_2}=R_{15},\\
\Phi_{0020}&=\phi_{0_10_20_10_2\dot{1}_1\dot{0}_2\dot{1}_1\dot{0}_2}=R_{55},\\
\Phi_{0110}&=\phi_{0_11_20_10_2\dot{1}_1\dot{0}_2\dot{0}_1\dot{0}_2}=R_{35},\\
\Phi_{1011}&=\phi_{1_10_20_10_2\dot{1}_1\dot{1}_2\dot{0}_1\dot{0}_2}=R_{14},\\
\Phi_{1101}&=\phi_{0_10_21_11_2\dot{0}_1\dot{0}_2\dot{0}_1\dot{1}_2}=R_{16},\\
\Phi_{2022}&=\phi_{1_10_21_10_2\dot{1}_1\dot{1}_2\dot{1}_1\dot{1}_2}=R_{44},\\
\Phi_{2202}&=\phi_{1_11_21_11_2\dot{0}_1\dot{1}_2\dot{0}_1\dot{1}_2}=R_{66},\\
\Phi_{1111}&=\phi_{1_11_20_10_2\dot{1}_1\dot{1}_2\dot{0}_1\dot{0}_2}=\dfrac{1}{3}(R_{12}+R_{34}+R_{56}),\\
\Phi_{1201}&=\phi_{1_10_21_11_2\dot{0}_1\dot{0}_2\dot{0}_1\dot{1}_2}=R_{36},\\
\Phi_{1021}&=\phi_{1_10_20_10_2\dot{1}_1\dot{1}_2\dot{1}_1\dot{0}_2}=R_{45},\\
\Phi_{2222}&=\phi_{1_11_21_11_2\dot{1}_1\dot{1}_2\dot{1}_1\dot{1}_2}=R_{22},\\
\Phi_{1211}&=\phi_{1_11_20_11_2\dot{1}_1\dot{1}_2\dot{0}_1\dot{0}_2}=R_{23},\\
\Phi_{2122}&=\phi_{1_11_21_10_2\dot{1}_1\dot{1}_2\dot{1}_1\dot{1}_2}=R_{24},\\
\Phi_{1121}&=\phi_{1_11_20_10_2\dot{1}_1\dot{1}_2\dot{1}_1\dot{0}_2}=R_{25},\\
\Phi_{2212}&=\phi_{1_11_21_11_2\dot{1}_1\dot{1}_2\dot{0}_1\dot{1}_2}=R_{26}.
\end{align}

\subsection{Commutation relation of the derivatives}
Covariant derivatives satisfy \cite{NP}:
\begin{equation}
\nabla_{\boldsymbol{a}\dot{\boldsymbol{b}}}\nabla_{\boldsymbol{c}\dot{\boldsymbol{d}}} - \nabla_{\boldsymbol{c}\dot{\boldsymbol{d}}}\nabla_{\boldsymbol{a}\dot{\boldsymbol{b}}} = \boldsymbol{\epsilon}^{\boldsymbol{pq}}\, \Big[ \Gamma_{\boldsymbol{pac}\dot{\boldsymbol{d}}}\partial_{\boldsymbol{q}\dot{\boldsymbol{b}}} - \Gamma_{\boldsymbol{pca}\dot{\boldsymbol{b}}}\partial_{\boldsymbol{q}\dot{\boldsymbol{d}}} \Big] + \boldsymbol{\epsilon}^{\dot{\boldsymbol{r}}\dot{\boldsymbol{s}}}\, \Big[ \overline{\Gamma}_{\dot{\boldsymbol{r}}\dot{\boldsymbol{b}}\dot{\boldsymbol{d}}\boldsymbol{c}}\,\partial_{\boldsymbol{a}\dot{\boldsymbol{s}}} - \overline{\Gamma}_{\dot{\boldsymbol{r}}\dot{\boldsymbol{d}}\dot{\boldsymbol{b}}\boldsymbol{a}}\,\partial_{\boldsymbol{c}\dot{\boldsymbol{s}}} \Big].
\end{equation}

Explicitely, we have:
\begin{align}
D\Delta-\Delta D    &= -\left(\bar{\alpha }_3+\alpha_3\right)\,\Delta +\left(\bar{\zeta }_3+\zeta_3\right)\,D+\alpha_4 \delta_2-\zeta_1\,\delta_1 ,\\
D\delta_1-\delta_1D &= \left(\alpha_2-\xi_1\right)\,\delta_1+(\xi_3-\alpha_4)\,D ,\\
D\delta_2-\delta_2D &= \left(\alpha_2-\bar{\alpha }_3\right)\,\delta_2 -\alpha_1 \,\Delta - \epsilon_1\,\delta_1+ \epsilon_3\,D,\\
D\delta_3-\delta_3D &= \left(\bar{\alpha }_2-\alpha_3\right)\,\delta_3-\beta_1\, \delta_1+\beta_3 \,D,\\
D\delta_4-\delta_4D &= \left(\bar{\alpha }_2-\alpha_3\right)\,\delta_4-\gamma_1\, \delta_1+\gamma_3\, D,\\
\Delta \delta_1-\delta_1\Delta &= \left(\zeta_2-\bar{\zeta }_3\right)\,\delta_1-\xi_4\,\delta_2-\zeta_4\,D+  \xi_3\,\Delta,\\
\Delta \delta_2-\delta_2\Delta &= -\left(\bar{\zeta }_3-\zeta_2+\epsilon_4\right)\,\delta_2+(\epsilon_3- \zeta_1)\,\Delta,\\
\Delta \delta_3-\delta_3\Delta &= \left(\bar{\zeta }_2-\zeta_3\right)\,\delta_3-\beta_4\, \delta_2+ \Delta\,\beta_3,\\
\Delta \delta_4-\delta_4\Delta &= \left(\bar{\zeta }_2-\zeta_3\right)\,\delta_4-\gamma_4 \,\delta_2+\gamma_3 \Delta,\\
\delta_1\delta_2-\delta_2\delta_1 &= \xi_2\,\delta_2-\epsilon_2\,\delta_1- \xi_1\,\Delta+ \epsilon_4\,D,\\
\delta_1\delta_3-\delta_3\delta_1 &= -\beta_2 \delta_1- \xi_3\,\delta_3+\beta_4 D,\\
\delta_1\delta_4-\delta_4\delta_1 &= -\gamma_2 \delta_1-\xi_3\,\delta_4 +\gamma_4 D,\\
\delta_2\delta_3-\delta_3\delta_2 &= -\beta_2 \delta_2+\beta_1 \Delta - \epsilon_3\delta_3,\\
\delta_2\delta_4-\delta_4\delta_2 &= -\gamma_2 \delta_2+\gamma_1 \Delta - \epsilon_3\delta_4,\\
\delta_3\delta_4-\delta_4\delta_3 &= \gamma_3 \delta_3-\beta_3 \delta_4.
\end{align}

\section{Expression of the vacuum Bianchi identities}
As we have found in Section \ref{sec:bia}, there are four sets of Bianchi identities:
\begin{align}
\nabla^{\boldsymbol{D}}_{\phantom{\boldsymbol{D}}\dot{\boldsymbol{G}}}\psi_{\boldsymbol{ABCD}}&=\nabla_{(\boldsymbol{C}}^{\phantom{\boldsymbol{C}}\dot{\boldsymbol{H}}}\phi_{\boldsymbol{AB})\dot{\boldsymbol{G}}\dot{\boldsymbol{H}}},\\
\nabla^{\boldsymbol{A}\dot{\boldsymbol{G}}}\phi_{\boldsymbol{AB}\dot{\boldsymbol{G}}\dot{\boldsymbol{H}}} &= -3\, \nabla_{\boldsymbol{B}\dot{\boldsymbol{H}}}\Lambda,\\
\nabla_{\boldsymbol{O}\dot{\boldsymbol{P}}}\,\chi_{\boldsymbol{A}\boldsymbol{C}\boldsymbol{I}\boldsymbol{M}} &=0,\\
\nabla_{\boldsymbol{O}\dot{\boldsymbol{P}}}\,\phi_{\boldsymbol{A}\boldsymbol{C}\dot{\boldsymbol{L}}\dot{\boldsymbol{N}}} &=0
\end{align}

Since their expression is long, we report them in the appendix \ref{app:bian}. Here we only write the vacuum expression of the first set, since we are going to use them later.

\begin{align}
D\,\psi_2-\delta_1\,\psi_0&= \alpha_4 \psi_0-\left(\alpha_2+\alpha_3\right) \psi_2+\alpha_1 \psi_5,\\
\Delta\,\psi_0-\delta_2\,\psi_1&=\psi_1 \left(-\zeta_1+\epsilon_2-\epsilon_3\right)-\psi_0 \left(\epsilon_4-2 \zeta_3\right),\\
D\,\psi_4-\delta_1\,\psi_1&= -\psi_1 \left(\alpha_4-2 \xi_2\right)+\psi_4 \left(-\alpha_2+\alpha_3-2 \xi_1\right)+\alpha_1 \psi_7,\\
\Delta\,\psi_1-\delta_2\,\psi_3&= \zeta_4 \psi_0+\psi_1 \left(-\zeta_2+\zeta_3-2 \epsilon_4\right)+\psi_3 \left(2 \epsilon_2-\zeta_1\right),\\
D\,\psi_5-\delta_1\,\psi_2&=\psi_2 \left(-2 \alpha_4+\xi_2-\xi_3\right)+\psi_5 \left(2 \alpha_3-\xi_1\right)+\psi_0 \epsilon_4,\\
\Delta\,\psi_5-\delta_2\,\psi_7&=-\psi_5 \left(\epsilon_4-2 \zeta_3\right)+\psi_7 \left(-2 \zeta_1+\epsilon_2-\epsilon_3\right)+\psi_6 \epsilon_1,\\
\Delta\,\psi_2-\delta_2\,\psi_4&=\psi_2 \left(\zeta_3-\epsilon_4\right)+\psi_4 \left(\epsilon_2-2 \zeta_1\right)+\psi_8 \epsilon_1,\\
D\,\psi_8-\delta_1\,\psi_3&=\psi_3 \left(\xi_2-\alpha_4\right)+\psi_8 \left(\alpha_3-\xi_1\right),\\
D \,\psi_6 -\delta_1\,\psi_8&= -\psi_8 \left(\alpha_4+\xi_3\right)+\alpha_3 \psi_6+\psi_3 \epsilon_4,\\
\Delta\,\psi_7-\delta_2\,\psi_6&=\psi_6 \left(-\zeta_1+2 \epsilon_2+\epsilon_4\right)-\psi_7 \left(\zeta_2-\zeta_3+\zeta_4+\epsilon_4\right),\\
D\,\psi_7-\delta_1\,\psi_4&=\psi_4 \left(-2 \alpha_4+\xi_2-\xi_3\right)+\psi_7 \left(2 \alpha_3-\xi_1\right)+\psi_1 \epsilon_4,\\
\Delta\,\psi_4-\delta_2\,\psi_8&=\psi_4 \left(\zeta_3-\epsilon_4\right)+\psi_8 \left(\epsilon_2-\zeta_1\right),\\
\delta_4\,\psi_0&=2 \gamma_3 \psi_0,\\
\delta_4\,\psi_1&=\gamma_4 \psi_0+\gamma_3 \psi_1,\\
\delta_3\,\psi_2&=\beta_3 \psi_2+\beta_1 \psi_5,\\
\delta_4\,\psi_2&=2\gamma_3 \psi_2,\\
\delta_3\,\psi_4&=\beta_3 \psi_4+\beta_1 \psi_7,\\
\delta_4\,\psi_4&=\gamma_4 \psi_2+\gamma_3 \psi_4,\\
\delta_3\,\psi_5&=2 \beta_3 \psi_5,\\
\delta_4\,\psi_5&=2 \gamma_3 \psi_5,\\
\delta_3\,\psi_6&=2 \beta_3 \psi_6,\\
\delta_3\,\psi_7&=2 \beta_3 \psi_7,\\
\delta_4\,\psi_7&=\gamma_4 \psi_5+\gamma_3 \psi_7,\\
\delta_3\,\psi_8&=\beta_1 \psi_6+\beta_3 \psi_8.
\end{align}

\section{The derivation of the higher dimensional analogue of the Teukolsky equation}
\label{sec:teueq}
In this Section, we procede to the calculation of the higher dimensional analogue of the Teukolsky equation. For simplicity, we still fix our spacetime to have 6 dimensions, the case of higher dimensional spacetime is analogous.

\subsection{6D Kerr metric}

We consider the following form of the 6d Kerr metric with the additional dimensions compactified to a two-sphere:
\begin{equation}
\begin{split}
ds^2&=\left( 1-\dfrac{2Mr}{\Sigma} \right)\,dt^2+\left( \dfrac{4Mar\,\sin^2\theta}{\Sigma} \right)dtd\phi - \dfrac{\Sigma}{\Delta}dr^2+\\
&-\Sigma\,d\theta^2-\sin^2\theta\left( r^2+a^2+\dfrac{2Ma^2r\sin^2\theta}{\Sigma} \right)d\phi^2-R^2\,(d\theta_1^2+\sin^2\theta_1d\theta_2^2)
\end{split}
\end{equation}
where, as usual:\footnote{Note that there is a clash in the notation: the $\Delta$ in this Section is not the derivative of the previous one.}
\begin{equation}
\Sigma=r^2+a^2\,\cos^2\theta, \qquad \Delta=r^2-2Mr+a^2,
\end{equation}
$M$ is the black hole mass, $aM$ is the angular momentum. Moreover, $\theta_1$ and $\theta_2$ are the angular variable of the compactified sphere and $R$ is its curvature radius.

\subsection{Vielbein}

The vielbein we consider is (compare to \cite{teu1}):
\begin{align}
l^\mu &= \left\{ \dfrac{a^2+r^2}{\Delta},1,0,\dfrac{a}{\Delta},0,0 \right\},\\
n^\mu &= \left\{ \dfrac{a^2+r^2}{2\Sigma},-\dfrac{\Delta}{2\Sigma},0,\dfrac{a}{2\Sigma},0,0 \right\},\\
m_1^\mu&=\left\{ ia\sin\theta,0,1,\dfrac{i}{\sin\theta},0,0 \right\}\,\dfrac{1}{\sqrt{2}\,(ia\cos\theta+r)},\\
m_1^\mu &= \overline{m}_2^\mu,\\
m_3^\mu &= \left\{ 0,0,0,0,-\dfrac{i}{\sqrt{2}\,R},-\dfrac{1}{\sqrt{2}\,R}\,\dfrac{1}{\sin\theta_1} \right\},\\
m_4^\mu &= \overline{m}_3^\mu.
\end{align}

\subsection{Spin coefficients}

The non-zero spin coefficients are:
\begin{align}
\alpha_4&=\dfrac{i\,a\,\sin\theta}{\sqrt{2}(a\cos\theta-ir)^2}\\
\beta_3&=-\dfrac{i\,\cot\theta_1}{2\sqrt{2}\,R}\\
\gamma_3&=-\dfrac{i\,\cot\theta_1}{2\sqrt{2}\,R}\\
\xi_1&=-\dfrac{1}{ia\cos\theta+r}\\
\xi_2&=-\dfrac{1}{\sqrt{2}\,\sin\theta}\,\dfrac{ia+r\cos\theta}{(a\cos\theta-ir)^2}\\
\xi_3&=-\dfrac{ia\sin\theta}{2\sqrt{2}(a\cos\theta-ir)^2}\\
\epsilon_2&=\dfrac{1}{\sqrt{2}\sin\theta}\,\dfrac{-ia+r\,\cos\theta}{(a\cos\theta+ir)^2}\\
\epsilon_3&=\dfrac{ia\sin\theta}{2\sqrt{2}\,(a\cos\theta+ir)^2}\\
\epsilon_4&=\dfrac{i\Delta}{2\Sigma\,(a\cos\theta-ir)}\\
\zeta_1&=\dfrac{ia\sin\theta}{\sqrt{2}\,\Sigma}\\
\zeta_2&=-\dfrac{ia\,\cos\theta\,\Delta}{\Sigma^2}\\
\zeta_3&=\dfrac{M\,(\Sigma-2r^2)}{2\,\Sigma^2}+\dfrac{a^2r\,\sin\theta}{2\,\Sigma^2}
\end{align}

\subsection{Weyl scalar}

The non-zero Weyl scalar are:
\begin{align}
\psi_3 &=\dfrac{1}{10R^2}+\dfrac{2Mr\,(4r^2-3\,\Sigma)}{\Sigma^2}\\
\psi_4 &=\dfrac{iaM\,\cos\theta\,(\Sigma-4r^2)}{\Sigma^2}\\
\psi_5 &=\dfrac{1}{30\,R^2}+\dfrac{2\,Mr\,(4r^2-3\,\Sigma)}{3\Sigma^2}
\end{align}

\subsection{The Ricci tensor}
{The considered metric does not have a zero Ricci tensor, in fact it is given by}
\begin{equation}
R_{\mu\nu}=\text{Diagonal}\left\{0,0,0,0,1,\sin^2(\theta_1)  \right\},
\end{equation}
{thus, the only non zero components are $R_{55}=\Phi_{0020}=1$ and $R_{66}={2202}=\sin^2(\theta_1)$. Luckily these components do not appear in the Bianchi identities we shall consider, in particular (B1) and (B2), thus, in the following, we can use the vacuum Bianchi identities, which we now write out.}

\subsection{The Bianchi identities}

The perturbed vacuum Bianchi identities are given by:
\begin{align}
&D\,\psi_2-(\delta_1-\alpha_4)\,\psi_0=\alpha_1\,\psi_5,\\
&(\Delta+\epsilon_4-2\zeta_3)\,\psi_0-(\delta_2+\zeta_1+\epsilon_3-\epsilon_2)\,\psi_1=0,\\
&(\alpha_4-2\xi_2-\delta_1)\,\psi_1=(\alpha_3-\alpha_2)\,\psi_4,\\
&(\Delta+2\epsilon_4+\zeta_2-\zeta_3)\,\psi_1=0,\\
&(\Delta+\epsilon_4-\zeta_3)\,\psi_2=0,\\
&(D+\xi_1-\epsilon_4)\,\psi_7=0,\\
&(\epsilon_3+2\zeta_1-\epsilon_2-\delta_2)\,\psi_7=0,\\
&(\zeta_1-2\epsilon_2-\delta_2)\,\psi_8=0,\\
&D\,\psi_6-(\delta_1+\alpha_4+\xi_3)=\xi_4\,\psi_3,\\
&(\Delta+\epsilon_4+\zeta_2-\zeta_3)\,\psi_7-(\delta_2+\zeta_1-2\epsilon_2-\epsilon_4)\,\psi_6=0,\\
&(D+\xi_1)\,\psi_8=0,\\
&(\delta_3-\beta_3)\,\psi_2=\beta_1\,\psi_5,\\
&(\delta_3-2\beta_3)\,\psi_6=0,\\
&(\delta_3-2\beta_3)\,\psi_7=0,\\
&(\delta_3-\beta_3)\,\psi_7=0,\\
&(\delta_4-2\gamma_3)\,\psi_0=0,\\
&(\delta_4-\gamma_3)\,\psi_1=0,\\
&(\delta_4-2\gamma_3)\,\psi_2=0,\\
&(\delta_4-\gamma_3)\,\psi_7=\gamma_4\,\psi_5.
\end{align}
{These equations were obtained by assuming that the spin coefficients and Weyl scalars have the form $a+da$, where $da$ is a perturbation. We have expanded the Bianchi identities keeping only terms at the first order in $da$. With an abuse of notation, we have indicated the perturbed spin coefficients and Weyl scalars with the same symbol as the unperturbed ones.}

The first order Bianchi identities are:
\begin{align}
&(D+2\xi_1)\,\psi_4=0,\\
&(\zeta_1-2\epsilon_2-\delta_2)\,\psi_3=0,\\
&(D-\xi_1)\,\psi_5=0,\\
&(2\zeta_1-\epsilon_2-\delta_2)\,\psi_4=0,\\
&(2\alpha_4+\xi_3-\xi_2-\delta_1)\,\psi_4=0,\\
&(\Delta+\epsilon_4-2\zeta_3)\,\psi_5=0,\\
&(\Delta+2\epsilon_4+\zeta_2-\zeta_3)\,\psi_4=0,\\
&(\alpha_4-\xi_2-\delta_1)\,\psi_3=0,\\
&(\delta_3-\beta_3)\,\psi_4=0,\\
&(\delta_3-2\beta_3)\,\psi_4=0,\\
&(\delta_4-\gamma_3)\,\psi_4=0,\\
&(\delta_4-2\gamma_3)\,\psi_5=0.
\end{align}
{These equations were obtained by a procedure similar to the above ones, keeping only terms at zero order in the perturbations.}

\subsection{Derivation of the Teukolsky equation}
We need to find an equation for $\psi_6$, since as we have seen above, it is related to the gravitational waves amplitude. We consider the following equations:
\begin{align}\label{eq:prima}
&D\,\psi_2-(\delta_1-\alpha_4)\,\psi_0=\alpha_1\,\psi_5,\\\label{eq:seconda}
&(\Delta+\epsilon_4-2\zeta_3)\,\psi_0-(\delta_2+\zeta_1+\epsilon_3-\epsilon_2)\,\psi_1=0,\\\label{eq:quarta}
&(\delta_4-2\gamma_3)\,\psi_0=0,\\\label{eq:uno}
&(D-\xi_1)\,\psi_5=0,\\\label{eq:ultima}
%(D+\xi_1)\,\psi_7&=0\\\label{eq:sesta}
%(\delta_2+\epsilon_2-\zeta_1-\epsilon_3)\,\psi_7&=0\\
%(\delta_3+\beta_3)\,\psi_7&=0\\
%(D-\xi_1)\,\psi_8&=0\\\label{eq:Bultima}
%(\delta_2+2\epsilon_2-\zeta_1)\,\psi_8&=0\\\label{eq:ultima}
&\left(\bar{\alpha }_3+\alpha_3+\xi_1-\alpha_1-\alpha_2-D\right)\epsilon_1+\left(\delta_2+\zeta_1-\epsilon_2-\epsilon_3\right)\alpha_1 =\psi_0.
\end{align}
We need to eliminate $\psi_1$ and $\psi_2$.

We first use \eqref{eq:uno} to rewrite \eqref{eq:ultima} as:
\begin{equation}\label{eq:ultima2}
\left(\bar{\alpha }_3+\alpha_3+2\xi_1-\alpha_1-\alpha_2-D\right)\epsilon_1\,\psi_5+\left(\delta_2+\zeta_1-\epsilon_2-\epsilon_3\right)\alpha_1\,\psi_5 =\psi_0\,\psi_5.
\end{equation}
We operate with $\left(\delta_2+\zeta_1-\epsilon_2-\epsilon_3\right)$ on \eqref{eq:prima} and with $\left(\bar{\alpha }_3+\alpha_3+2\xi_1-\alpha_1-\alpha_2-D\right)$ on \eqref{eq:seconda}. The terms in $\psi_1$ and $\psi_2$ simplify due to the commutation relations and the derivatives of the spin coefficients. The remaining terms can be simplified with \eqref{eq:ultima2}, so we find:
\begin{equation}\label{eq:parte}
\left(\bar{\alpha }_3+\alpha_3+2\xi_1-\alpha_1-\alpha_2-D\right)(\Delta+\epsilon_4-2\zeta_3)\,\psi_0-\left(\delta_2+\zeta_1-\epsilon_2-\epsilon_3\right)(\delta_1-\alpha_4)\,\psi_0-\psi_5\,\psi_0=0
\end{equation}

We derive \eqref{eq:quarta} by $\delta_3$ on the left, thus obtaining
\begin{equation}
\delta_3(\delta_4-2\gamma_3)\,\psi_0=0
\end{equation}
and subtract this expression from \eqref{eq:parte}, thus obtaining:
\begin{equation}
\begin{split}
&\left(\bar{\alpha }_3+\alpha_3+2\xi_1-\alpha_1-\alpha_2-D\right)(\Delta+\epsilon_4-2\zeta_3)\,\psi_0-\left(\delta_2+\zeta_1-\epsilon_2-\epsilon_3\right)(\delta_1-\alpha_4)\,\psi_0+\\
&-\delta_3(\delta_4-2\gamma_3)\,\psi_0-\psi_5\,\psi_0=0
\end{split}
\end{equation}
This is the decoupled equation for $\psi_0$ we were looking for. Then, as in \cite{teu1} we rotate the vielbein to obtain an expression for $\psi_6$:
\begin{equation}\label{eq:parte2}
\begin{split}
&\left(\bar{\zeta}_3+\zeta_3-2\zeta_2+\zeta_4+\epsilon_4-\Delta\right)(D+\xi_1-2\alpha_3)\,\psi_6-\left(\delta_1+\alpha_4-\xi_2-\xi_3\right)(\delta_2-\zeta_1)\,\psi_6+\\
&-\delta_4(\delta_3-2\beta_3)\,\psi_6-\psi_5\,\psi_6=0.
\end{split}
\end{equation}

Substituting the expressions of the derivatives and of the spin coefficients, we obtain imposing
\begin{equation}
\psi=\dfrac{\psi_6}{(r-ia\cos\theta)^4}
\end{equation}
\begin{equation}
\begin{split}
&\left( \dfrac{(a^2+r^2)^2}{\Delta} - a^2\sin^2\theta \right)\, \dfrac{\partial^2\psi}{\partial t^2} + \dfrac{4Ma r}{\Delta}\,\dfrac{\partial^2\psi}{\partial t\partial \phi}+\left( \dfrac{a^2}{\Delta}-\dfrac{1}{\sin^2\theta} \right)\,\dfrac{\partial^2\psi}{\partial\phi^2} -\Delta\,\dfrac{\partial^2\psi}{\partial r^2} +\\
-& 6(r-M)\,\dfrac{\partial\psi}{\partial r}-\dfrac{\partial^2\psi}{\partial \theta^2} -\cot\theta \dfrac{\partial\psi}{\partial\theta}-4\left( \dfrac{a(r-M)}{\Delta}+\dfrac{i\cos\theta}{\sin^2\theta} \right)\,\dfrac{\partial\psi}{\partial\phi}+\\
-&4\left( \dfrac{M(r^2-a^2)}{\Delta}-r-ia\cos\theta \right)\,\dfrac{\partial\psi}{\partial t}+ \left( 4\cot^2\theta - 2 + \dfrac{2}{15R^2} \right)\,\psi +\\
+&\dfrac{1}{2R^2}\,\left[ -\dfrac{\partial\psi}{\partial\theta_1^2}-\dfrac{1}{\sin^2\theta_1}\,\dfrac{\partial\psi}{\partial\theta_2^2}-\cot\theta_1\dfrac{\partial\psi}{\partial\theta_1}-i\dfrac{\cos\theta_1}{\sin^2\theta_1}\,\dfrac{\partial\psi}{\partial\theta_2} + \dfrac{1}{\sin^2\theta_1}\,\psi\right]=0
\end{split}
\end{equation}

%%%%%%%%%%%%%%%%%%%%%%%%%%%%%%%%%%%%%%%%%%%%%%%%%%%%%%%%%%%%%%%%%%%%%

We first notice that the last line separates from the rest, so we introduce $\psi=F(t,r,\theta,\phi)\Theta(\theta_1\theta_2)$. The part in the square brackets solves the equation
\begin{equation}\label{eq:new}
-\dfrac{\partial\Theta}{\partial\theta_1^2}-\dfrac{1}{\sin^2\theta_1}\,\dfrac{\partial\Theta}{\partial\theta_2^2}-\cot\theta_1\dfrac{\partial\Theta}{\partial\theta_1}-i\dfrac{\cos\theta_1}{\sin^2\theta_1}\,\dfrac{\partial\Theta}{\partial\theta_2} + \dfrac{1}{\sin^2\theta_1}\,\Theta = -L(L+1) \,\Theta,
\end{equation}
where $L(L+1)$ is a separation constant.

Due to the similarity between the above equation and the spherical harmonic equation, we write \cite{huges}:
\begin{equation}
\Theta(\theta_1\theta_2)=\sum_{lm}b_{lm}Y_{lm}(\theta_1,\theta_2)
\end{equation}
Substituting, we obtain:
\begin{equation}
\sum b_{lm}\left\{-2i\dfrac{\cot\theta_1}{\sin^2\theta_1}\,\dfrac{\partial\,Y_{lm}}{\partial\theta_2}+\dfrac{1}{\sin^2\theta_1}\,Y_{lm}+\Big[ L(L+1)+l(l+1) \Big]\,Y_{lm} \right\}=0,
\end{equation}
where $Y_{lm}(\theta_1,\theta_2)$ are the spherical harmonics. For higher dimensional spacetimes, we need to use the function reported in \cite{aguci} in place of the spherical harmonics, but for the rest the procedure is analogous. We  now multiply the above expression on the left by $\overline{Y}_{l_1m_1}(\theta_1,\theta_2)$ and integrate. The result is a band matrix, which can be diagonalized, so that the eigenvalues $L(L+1)$ can be obtained. The first few eigenvalues are given in table \ref{tab:eigen}.
\begin{table}[ht]
\centering
\begin{tabular}{cc|cc}
(l,m) & L       & (l,m)   & L\\
\hline
(2,2) & 2.38954 & (7,7)   & 7.19806\\
(3,3) & 3.29710 & (8,8)   & 8.18886\\
(4,4) & 4.25375 & (9,9)   & 9.18172\\
(5,5) & 5.22764 & (10,10) & 10.1760\\
(6,6) & 6.21035 & (11,11) & 11.1714\\
\end{tabular}
\caption{The first eigenvalues of equation \eqref{eq:new}}\label{tab:eigen}
\end{table}

The rest of the equation now reads:
\begin{equation}
\begin{split}
&\left[ \dfrac{(a^2+r^2)^2}{\Delta} - a^2\sin^2\theta \right]\, \dfrac{\partial^2F}{\partial t^2} + \dfrac{4Ma r}{\Delta}\,\dfrac{\partial^2F}{\partial t\partial \phi}+\left[ \dfrac{a^2}{\Delta}-\dfrac{1}{\sin^2\theta} \right]\,\dfrac{\partial^2F}{\partial\phi^2} -\Delta\,\dfrac{\partial^2F}{\partial r^2}+\\
-&6(r-M)\,\dfrac{\partial F}{\partial r}-\dfrac{\partial^2F}{\partial \theta^2} -\cot\theta \dfrac{\partial F}{\partial\theta}-4\left[ \dfrac{a(r-M)}{\Delta}+\dfrac{i\cos\theta}{\sin^2\theta} \right]\,\dfrac{\partial F}{\partial\phi}+\\
-&4\left[ \dfrac{M(r^2-a^2)}{\Delta}-r-ia\cos\theta \right]\,\dfrac{\partial F}{\partial t}+ \left( 4\cot^2\theta - 2 + \dfrac{2}{15R^2} + \dfrac{1}{2R^2}L(L+1) \right)\,F =0
\end{split}
\end{equation}

As usual, this can be furhter separated into a radial and an angular equation defining:
\begin{equation}
F=\exp(-i\omega t)\,\exp(im \phi) \,T_{lm}(\theta)\,R_{lmL}(r).
\end{equation}
The angular equation is the usual spheroidal harmonics equation; the radial one is similar to that of Teukolsky,  but there is one more term in the potential depending on the radius of the compactified metric and on the eigenvalues of equation \eqref{eq:new}. Using Teukolsky notation \cite{teu1}, we find:
\begin{equation}
\begin{split}
&\Delta\,\dfrac{\partial^2R_{lmL}(r)}{\partial r^2} + 6(r-M)\,\dfrac{\partial R_{lmL}(r)}{\partial r} +\\
+& \left[ \dfrac{K^2-4i(r-M)K}{\Delta}+8i\omega\,r-a^2\omega^2+2am\omega-A + \dfrac{4+15L(L+1)}{30R^2} \right]\,R_{lmL}(r)=0,
\end{split}
\end{equation}
{where}
\begin{equation}
    K=(r^2+a^2)\,\omega-a m
\end{equation}
{and $A$ is a separation constant to be calculated. We notice that this equation coincides with that reported in \cite{highKerr1,highKerr2,highKerr3}.}

Notice that the radial function has one more index $L$, the eigenvalue of the additional dimension while $l$ and $m$ have the usual meaning \cite{teu1}.

{The boundary conditions for the above equation are: no incoming radiation from infinity, only outgoing waves.}

\section{Quasi-normal modes}
\label{sec:QNM}

In order to find the QNM, we follow \cite{leaver} and use the continued fraction method.

Since the equation is very similar to Teukolsky's, the procedure is also very similar. We define\footnote{We also fix $M=1/2$.} $\Delta=(r-r_{+})(r-r_{-})$, $b=\sqrt{1-4a^2}$ and $\sigma_{+}=(\omega r_{+}-am)/b$ and introduce the series representation:
\begin{equation}
R_{lmL}(r)=\exp(i\omega r)(r-r_{-})^{-3+i\omega+i\sigma_{+}}(r-r_{+})^{-2-i\sigma_{+}}\sum d_n\left( \dfrac{r-r_{+}}{r-{r_{-}}} \right)^n
\end{equation}
which permits us to find the recursion relation:
\begin{align}
&\alpha^r_0\,d_1+\beta^r_0\,d_0=0\\
&\alpha^r_n\,d_{n+1}+\beta^r_n\,d_n+\gamma^r_{n}\,d_{n-1}=0
\end{align}
with 
\begin{align}
\alpha^r_n&=n^2+(c_0+1)+c_0\\
\beta^r_n &=-2n^2+(c_1+2)n+c_3\\
\gamma^r_n&=n^2+(c_2-3)n+c_4-c_2+2\\
c_0&=3-i\omega-\dfrac{2i}{b}\,\left( \dfrac{\omega}{2}-am \right)\\
c_1&=-4+2i\omega(2+b)+\dfrac{4i}{b}\,\left( \dfrac{\omega}{2}-am \right)\\
c_2&=1-i\omega-\dfrac{2i}{b}\,\left( \dfrac{\omega}{2}-am \right)\\
c_3&=\omega^2(4+2b-a^2)-2am\omega+1+(2+b)i\omega-A+\dfrac{4\omega+2i}{b}\left( \dfrac{\omega}{2}-am \right)+\dfrac{4+15L(L+1)}{30R^2}\\
c_4&=-1-2\omega^2+i\omega-\dfrac{4\omega+2i}{b}\,\left( \dfrac{\omega}{2}-am \right)
\end{align}

For the angular equation, we have the recursion relation:
\begin{align}
&\alpha^\theta_0\,d_1+\beta^\theta_0\,d_0=0\\
&\alpha^\theta_n\,d_{n+1}+\beta^\theta_n\,d_n+\gamma^\theta_{n}\,d_{n-1}=0
\end{align}
where
\begin{align}
\alpha^\theta_n&=-2(n+1)(n+2k_1+1)\\
\beta^\theta_n&=n(n-1)+2n(k_1+k_2+1-a\omega)-\Big[ 2a\omega(2k_1-1)-(k_1+k_2)(k_1+k_2+1) \Big] - \Big[ a^2\omega^2+2-A \Big]\\
\gamma^\theta_n&=2a\omega(n+k_1+k_2-2)
\end{align}
with $k_1=|m-s|/2$, $k_2=|m+s|/2$.

In tables \ref{tab:QNM} -- \ref{tab:QNM_ultima} we report the fundamental mode for various values of the parameters. For comparison, in table \ref{tab:comp} we report the fundamental mode (as calculated in \cite{leaver}) for the 4d Kerr black hole for the same values of $a$ used in the other tables. {We have truncated the continued fraction at sixth order and solved the equation with the Mathematica built-in function NSolve.}

{By comparing the QNM calculated in this paper, with those of the 4d Kerr black hole in table \ref{tab:comp}, we notice that the QNM obtained here are always smaller than the 4d ones; we also notice that the larger the value of $R$, the closer they are to the 4d case. Moreover, as $L$ grows, the real and imaginary parts of the QNM decrease. The effect of $L$ is larger for smaller values of $R$.} {We also notice that when the parameters $R$ and $L$ are fixed, the real parts of the QNM increase with increasing the parameter $a$, similar to the 4d results. On the other hand, when $R$ and $L$ are fixed, the imaginary parts of the QNM increase with increasing $a$ for $a<0.2$ and decrease for $a>0.2$. This is in contrast to the usual 4d case and, in principle, can be used to verify the presence of compactified extra dimensions.}

{QNM for higher dimensional Kerr black hole were calculated in the already cited \cite{highKerr1,highKerr2,highKerr3}: the first derives analytic expressions for the QNM in $d=\{4,5,7\}$, while the other use a numerical approach to derive the QNM for scalar perturbations. It would be interesting to compare our results with \cite{highKerr1}, however, it is not possible, since they calculate QNM for $d=\{4,5,7\}$. We tried to derive from their formulae the case $d=6$, but, unfortunately, we could not find an analytic expression for the various functions, thus for $d=6$ their method is not applicable; on the other side, as we said above, our method works only for even dimensional spacetime, so a direct comparison with that paper is not possible.}

{Moreover, we notice that our procedure can be extended to calculate the overtones which are important in the waveform soon after the merger, thus our work could also be used to study black holes in this respect.}

\begin{table}[h]
\centering
\begin{tabular}{cccc}
R    & L       & A & $\omega$\\
\hline
     & 2.38954 & 4 & 0.702014 - 0.100732 i\\
     & 3.29710 & 4 & 0.702013 - 0.100732 i\\
1000 & 4.25375 & 4 & 0.702013 - 0.100732 i\\
     & 5.22764 & 4 & 0.702013 - 0.100732 i\\
     & 6.21035 & 4 & 0.702012 - 0.100732 i\\
\hline
     & 2.38954 & 4 & 0.701973 - 0.100728 i\\
     & 3.29710 & 4 & 0.701944 - 0.100726 i\\
100  & 4.25375 & 4 & 0.701904 - 0.100722 i\\
     & 5.22764 & 4 & 0.701854 - 0.100718 i\\
     & 6.21035 & 4 & 0.701794 - 0.100713 i\\
\hline
     & 2.38954 & 4 & 0.697924 - 0.100381 i\\
     & 3.29710 & 4 & 0.694942 - 0.100123 i\\
10   & 4.25375 & 4 & 0.690912 - 0.099773 i\\
     & 5.22764 & 4 & 0.685853 - 0.099328 i\\
     & 6.21035 & 4 & 0.679755 - 0.0987\,10\,6 i\\
\end{tabular}
\caption{Fundamental mode for the six-dimensional Kerr black hole for $a=0$.}\label{tab:QNM}
\end{table}

\begin{table}[h]
\centering
\begin{tabular}{cccc}
R    & L       &         A            & $\omega$\\
\hline
     & 2.38954 & 4.06842 - 0.0093511 i & 0.711188 - 0.101135 i\\
     & 3.29710 & 4.06842 - 0.0093511 i & 0.711188 - 0.101135 i\\
1000 & 4.25375 & 4.06842 - 0.0093511 i & 0.711187 - 0.101135 i\\
     & 5.22764 & 4.06842 - 0.0093511 i & 0.711187 - 0.101135 i\\
     & 6.21035 & 4.06842 - 0.0093511 i & 0.711186 - 0.101135 i\\
\hline
     & 2.38954 & 4.06842 - 0.0093528 i & 0.711148 - 0.101132 i\\
     & 3.29710 & 4.06842 - 0.0093524 i & 0.711118 - 0.101129 i\\
100  & 4.25375 & 4.06842 - 0.0093524 i & 0.71107\,10\, - 0.101126 i\\
     & 5.22764 & 4.06842 - 0.0093520 i & 0.711028 - 0.101122 i\\
     & 6.21035 & 4.06842 - 0.0093516 i & 0.710968 - 0.101117 i\\
\hline
     & 2.38954 & 4.06805 - 0.0093254 i & 0.707084 - 0.100794 i\\
     & 3.29710 & 4.06777 - 0.0093050 i & 0.704091 - 0.100543 i\\
10   & 4.25375 & 4.06739 - 0.0092772 i & 0.700048 - 0.100201 i\\
     & 5.22764 & 4.06692 - 0.0092418 i & 0.694971 - 0.099769 i\\
     & 6.21035 & 4.06636 - 0.0091986 i & 0.688851 - 0.099241 i\\
\end{tabular}
\caption{Fundamental mode for the six-dimensional Kerr black hole for $a=0.2$.}
\end{table}

\begin{table}[h]
\centering
\begin{tabular}{cccc}
R    & L       &         A            & $\omega$\\
\hline
     & 2.38954 & 4.20650 - 0.0277263 i & 0.745965 - 0.093157 i\\
     & 3.29710 & 4.20650 - 0.0277263 i & 0.745965 - 0.093157 i\\
1000 & 4.25375 & 4.20650 - 0.0277263 i & 0.745964 - 0.093157 i\\
     & 5.22764 & 4.20650 - 0.0277263 i & 0.745964 - 0.093157 i\\
     & 6.21035 & 4.20649 - 0.0277262 i & 0.745963 - 0.093157 i\\
\hline
     & 2.38954 & 4.20648 - 0.0277224 i & 0.745924 - 0.093154 i\\
     & 3.29710 & 4.20647 - 0.0277223 i & 0.745893 - 0.093152 i\\
100  & 4.25375 & 4.20646 - 0.0277221 i & 0.745852 - 0.093149 i\\
     & 5.22764 & 4.20645 - 0.0277186 i & 0.745802 - 0.093145 i\\
     & 6.21035 & 4.20643 - 0.0277157 i & 0.745740 - 0.093141 i\\
\hline
     & 2.38954 & 4.20528 - 0.0275287 i & 0.74177\,10\, - 0.092846 i\\
     & 3.29710 & 4.20439 - 0.0273862 i & 0.738725 - 0.092618 i\\
10   & 4.25375 & 4.20320 - 0.0271956 i & 0.734600 - 0.092308 i\\
     & 5.22764 & 4.20171 - 0.0269595 i & 0.729421 - 0.091914 i\\
     & 6.21035 & 4.19992 - 0.0266794 i & 0.72317\,10\, - 0.091434 i\\
\end{tabular}
\caption{Fundamental mode for the six-dimensional Kerr black hole for $a=0.6$.}
\end{table}

\begin{table}[h]
\centering
\begin{tabular}{cccc}
R    & L       &         A            & $\omega$\\
\hline
     & 2.38954 & 4.30534 - 0.0504171 i & 0.771122 - 0.081232 i\\
     & 3.29710 & 4.30534 - 0.0504170 i & 0.771121 - 0.081232 i\\
1000 & 4.25375 & 4.30534 - 0.0504169 i & 0.771121 - 0.081232 i\\
     & 5.22764 & 4.30534 - 0.0504168 i & 0.771120 - 0.081231 i\\
     & 6.21035 & 4.30534 - 0.0504167 i & 0.771120 - 0.081231 i\\
\hline
     & 2.38954 & 4.30532 - 0.0504086 i & 0.771079 - 0.081229 i\\
     & 3.29710 & 4.30530 - 0.0504025 i & 0.771048 - 0.081227 i\\
100  & 4.25375 & 4.30527 - 0.0503941 i & 0.771006 - 0.081224 i\\
     & 5.22764 & 4.30524 - 0.0503837 i & 0.770953 - 0.081222 i\\
     & 6.21035 & 4.30521 - 0.0503713 i & 0.770891 - 0.081215 i\\
\hline
     & 2.38954 & 4.30284 - 0.0495728 i & 0.766816 - 0.080911 i\\
     & 3.29710 & 4.30104 - 0.0489703 i & 0.763677 - 0.080677 i\\
10   & 4.25375 & 4.29863 - 0.0481734 i & 0.759438 - 0.080359 i\\
     & 5.22764 & 4.29565 - 0.0472000 i & 0.754118 - 0.079957 i\\
     & 6.21035 & 4.29211 - 0.0460654 i & 0.747709 - 0.079470 i\\
\end{tabular}
\caption{Fundamental mode for the six-dimensional Kerr black hole for $a=0.8$.}\label{tab:QNM_ultima}
\end{table}

\begin{table}[h]
\centering
\begin{tabular}{ccc}
a    &         A            & $\omega$\\
\hline
0    & 4                    & 0.747343 - 0.177925 i\\
0.2  & 3.99722 + 0.00139 i  & 0.750248 - 0.177401 i\\
0.6  & 3.97297 + 0.01262 i  & 0.776108 - 0.171989 i\\
0.8  & 3.94800 + 0.02226 i  & 0.803835 - 0.164313 i\\
\end{tabular}
\caption{Fundamental QNM for the four-dimensional Kerr black hole for various values of $a$. Taken from \cite{leaver}.}\label{tab:comp}
\end{table}

\section{Conclusion}
\label{sec:concl}
We have presented a method based on vectors of two-spinors (we have called them multispinors) with which we have derived a formalism analogue to that of Newman-Penrose, but applied to higher dimensional spacetimes. With this formalism we have derived the higher dimensional analogue of the Teukolsky equation for gravitational perturbations of rotating Kerr black hole with additional dimension compactified to an hypersphere. With this we have calculated with the usual means the QNM for different values of the curvature radius of the hypersphere.

We have also shown that with the method of multispinors we can reproduce the higher dimensional Lorentz group, in a similar way to what is done with spinors in 4d.

In a forthcoming paper, we shall show how to use the miltispinor formalism to derive a classification of higher dimensional spacetimes similar to that of \cite{hdc1,hdc2}.

\appendix
\section{Derivatives of spin coefficients}
\label{app:der}
The derivatives of the spin coefficients are:
\begin{align}
&\left(-\bar{\zeta }_3+\Delta -\zeta_2-2 \zeta_3\right)\,\alpha_1+ \left(\bar{\alpha }_3+\alpha_2+2 \alpha_3-D+\xi_1\right)\,\zeta_1-\Phi_{1011}-\alpha_4 \epsilon_1-\psi_2=0,\\
&\left(\bar{\alpha }_3+\alpha_3-D+\xi_1\right)\,\xi_1 +\left(\alpha_4+\delta_1-\xi_2-2 \xi_3\right)\,\alpha_1 =0,\\
&\left(\bar{\alpha }_3+\alpha_3-D+\xi_1\right)\,\epsilon_1 +\left(\delta_2+\zeta_1-\epsilon_2-2 \epsilon_3\right)\,\alpha_1 -\psi_0=0,\\
&\left(-\bar{\alpha }_2+\alpha_2+2 \alpha_3-D+\xi_1\right)\,\beta_1 +\left(-\beta_2-2 \beta_3+\delta_3\right)\,\alpha_1 =0,\\
&\left(-\bar{\alpha }_2+\alpha_2+2 \alpha_3-D+\xi_1\right)\,\gamma_1 +\left(-\gamma_2-2 \gamma_3+\delta_4\right)\,=0,\\
%%%%%%%%%%%%
&\left(-\bar{\zeta }_3+\Delta -\zeta_3\right)\,\alpha_2 +\left(\bar{\alpha }_3+\alpha_3-D\right)\,\zeta_2 -\alpha_1 \zeta_4+\alpha_4 \left(\zeta_1-\epsilon_2\right)+\zeta_1 \xi_2-\psi_4=0,\\
&\left(\bar{\alpha }_3-D+\xi_1\right)\,\xi_2 +\left(\alpha_4+\delta_1-\xi_2-\xi_3\right)\,\alpha_2 +\alpha_4 \xi_1-\alpha_1 \xi_4=0,\\
&\left(\bar{\alpha }_3-D\right)\,\epsilon_2 +\left(\delta_2-\epsilon_2-\epsilon_3\right)\,\alpha_2 +\alpha_1 \left(\zeta_2-\epsilon_4\right)+\epsilon_1 \left(\alpha_4+\xi_2\right)-\psi_1=0,\\
&-\left(\bar{\alpha }_2-\alpha_3+D\right)\,\beta_2 + \left(\delta_3-\beta_3\right)\,\beta_1+ \left(\alpha_4+\xi_2\right)\,\alpha_2-\alpha_1 \beta_4=0,\\
&-\left(\bar{\alpha }_2-\alpha_3+D\right)\,\gamma_2 +\left(\delta_4-\gamma_3\right)\,\alpha_2 +\gamma_1 \left(\alpha_4+\xi_2\right)-\alpha_1 \gamma_4=0,\\
%%%%%%%%%%%%
&\left(\Delta -\bar{\zeta }_3\right)\,\alpha_3 +\left(\bar{\alpha }_3-D\right)\,\zeta_3 -\Phi_{1111}-\alpha_1 \zeta_4+\alpha_4 \left(\zeta_1-\epsilon_3\right)+\zeta_1 \xi_3-\Lambda -\psi_5=0,\\
&\left(\bar{\alpha }_3-\alpha_2-D+\xi_1\right)\,\xi_3 +\left(\alpha_4+\delta_1-\xi_3\right)\,\alpha_3 +\alpha_4 \xi_1-\alpha_1 \xi_4=0,\\
&-\left(-\bar{\alpha }_3+\alpha_2+D\right)\,\epsilon_3 +\left(\delta_2-\epsilon_3\right)\,\alpha_3 +\alpha_1 \left(\zeta_3-\epsilon_4\right)+\epsilon_1 \left(\alpha_4+\xi_3\right)-\psi_2=0,\\
&-\left(\bar{\alpha }_2+D\right)\,\beta_3 +\beta_1 \left(\alpha_4+\xi_3\right)-\alpha_1 \beta_4+ \delta_3\,\alpha_3=0,\\
&-\left(\bar{\alpha }_2+D\right)\,\gamma_3 -\Phi_{1101}+\gamma_1 \left(\alpha_4+\xi_3\right)-\alpha_1 \gamma_4+ \delta_4\alpha_3=0,\\
%%%%%%%%%%%%
&\left(-\bar{\zeta }_3+\Delta +\zeta_2-\epsilon_4\right)\,\alpha_4 -\left(-\bar{\alpha }_3+\alpha_2+D\right)\,\zeta_4 -\Phi_{1211}+\zeta_1 \xi_4-\psi_7=0,\\
&\left(\bar{\alpha }_3-2 \alpha_2-\alpha_3-D+\xi_1\right)\,\xi_4 +\left(\alpha_4+\delta_1+\xi_2\right)\,\alpha_4 =0,\\
&-\left(-\bar{\alpha }_3+2 \alpha_2+\alpha_3+D\right)\,\epsilon_4 +\left(\delta_2+\epsilon_2\right)\,\alpha_4 +\alpha_1 \zeta_4+\Lambda -\psi_4+\xi_4 \epsilon_1=0,\\
&-\left(\bar{\alpha }_2+\alpha_2+D\right)\,\beta_4 +\left(\beta_2+\delta_3\right)\,\alpha_4 +\beta_1 \xi_4=0,\\
&-\left(\bar{\alpha }_2+\alpha_2+D\right)\,\gamma_4 -\Phi_{1201}+ \left(\gamma_2+\delta_4\right)\,\alpha_4+\gamma_1 \xi_4=0,\\
%%%%%%%%%%%%%%%%%%%%%%%%
%%%%%%%%%%%%
&\left(\bar{\zeta }_2+\Delta -\zeta_2-2 \zeta_3\right)\,\beta_1 +\left(\beta_2+2 \beta_3-\delta_3\right)\,\zeta_1 -\beta_4 \epsilon_1=0,\\
&\alpha_1 \beta_4+\left(\beta_3-\delta_3\right)\,\xi_1 +\left(\delta_1-\xi_2-2 \xi_3\right)\,\beta_1 =0,\\
&\left(\delta_2+\zeta_1-\epsilon_2-2 \epsilon_3\right)\,\beta_1 +\left(\beta_3-\delta_3\right)\,\epsilon_1 =0,\\
&\left(\beta_2+2 \beta_3-\delta_3\right)\,\gamma_1 +\left(-\gamma_2-2 \gamma_3+\delta_4\right)\,\beta_1 +\psi_2=0,\\
%%%%%%%%%%%%
&\left(\bar{\zeta }_2+\Delta -\zeta_3\right)\,\beta_2 +\left(\beta_3-\delta_3\right)\,\zeta_2 -\beta_1 \zeta_4+\beta_4 \left(\zeta_1-\epsilon_2\right)=0,\\
&\beta_4 \left(\alpha_2+\xi_1\right)+ \left(\delta_1-\xi_2-\xi_3\right)\,\beta_2-\beta_1 \xi_4-\delta_3 \xi_2=0,\\
&\left(\delta_2-\epsilon_2-\epsilon_3\right)\,\beta_2 +\beta_1 \left(\zeta_2-\epsilon_4\right)+\beta_4 \epsilon_1-\delta_3 \epsilon_2=0,\\
&\left(\beta_3-\delta_3\right)\,\gamma_2 +\left(\delta_4-\gamma_3\right)\,\beta_2 +\beta_4 \gamma_1-\beta_1 \gamma_4+\psi_4=0,\\
%%%%%%%%%%%%
&\left(\bar{\zeta }_2+\Delta \right)\,\beta_3 -\Phi_{1121}+\beta_1 \zeta_4- \left(\zeta_1+\epsilon_3\right)\,\beta_4-\delta_3 \zeta_3=0,\\
&\beta_4 \left(\alpha_3-\xi_1\right)+ \left(\delta_1-\xi_3\right)\,\beta_3- \left(\beta_2+\delta_3\right)\,\xi_3+\beta_1 \xi_4=0,\\
&\left(\delta_2-\epsilon_3\right)\,\beta_3 -\left(\beta_2+\delta_3\right)\,\epsilon_3 +\beta_1 \left(\zeta_3+\epsilon_4\right)-\beta_4 \epsilon_1=0,\\
&-\beta_4 \gamma_1+\beta_1 \gamma_4+ \delta_4\,\beta_3- \delta_3\,\gamma_3+\Lambda +\psi_5=0,\\
%%%%%%%%%%%%
&\left(\bar{\zeta }_2+\Delta -\zeta_2-2 \zeta_3-\epsilon_4\right)\,\beta_4 + \left(\beta_2+2 \beta_3-\delta_3\right)\,\zeta_4=0,\\
&\left(\alpha_4+\delta_1-\xi_2-2 \xi_3\right)\,\beta_4 +\left(\beta_3-\delta_3\right)\,\xi_4 =0,\\
&\left(\delta_2-\epsilon_2-2 \epsilon_3\right)\,\beta_4 +\left(\beta_3-\delta_3\right)\,\epsilon_4 +\beta_1 \zeta_4=0,\\
&\left(\beta_2+2 \beta_3-\delta_3\right)\,\gamma_4 +\left(-\gamma_2-2 \gamma_3+\delta_4\right)\,\beta_4 +\psi_7=0,\\
%%%%%%%%%%%%%%%%%%%%%%%%
%%%%%%%%%%%%
&\left(\bar{\zeta }_2+\Delta -\zeta_2-2 \zeta_3\right)\,\gamma_1 +\left(\gamma_2+2 \gamma_3-\delta_4\right)\,\zeta_1 -\gamma_4 \epsilon_1=0,\\
&\alpha_1 \gamma_4+\left(\gamma_3-\delta_4\right)\,\xi_1 +\left(\delta_1-\xi_2-2 \xi_3\right)\,\gamma_1 =0,\\
&\left(\delta_2+\zeta_1-\epsilon_2-2 \epsilon_3\right)\,\gamma_1 +\left(\gamma_3-\delta_4\right)\,\epsilon_1 =0,\\
%%%%%%%%%%%%
&\left(\bar{\zeta }_2+\Delta -\zeta_3\right)\,\gamma_2 +\left(\gamma_3-\delta_4\right)\,\zeta_2 -\gamma_1 \zeta_4+\gamma_4 \left(\zeta_1-\epsilon_2\right)=0,\\
&\gamma_4 \left(\alpha_2+\xi_1\right)+\left(\delta_1-\xi_2-\xi_3\right)\,\gamma_2 -\gamma_1 \xi_4-\delta_4 \xi_2=0,\\
&\left(\delta_2-\epsilon_2-\epsilon_3\right)\,\gamma_2 +\gamma_1 \left(\zeta_2-\epsilon_4\right)+\gamma_4 \epsilon_1-\delta_4 \epsilon_2=0,\\
%%%%%%%%%%%%
&\left(\bar{\zeta }_2+\Delta \right)\,\gamma_3 -\gamma_1 \zeta_4+\gamma_4 \left(\zeta_1-\epsilon_3\right)-\delta_4 \zeta_3=0,\\
&\gamma_4 \left(\alpha_3+\xi_1\right)+\left(\delta_1-\xi_3\right)\,\gamma_3 -\left(\gamma_2+\delta_4\right)\,\xi_3 -\gamma_1 \xi_4=0,\\
&\left(\delta_2-\epsilon_3\right)\,\gamma_3 -\left(\gamma_2+\delta_4\right)\,\epsilon_3 +\gamma_1 \left(\zeta_3-\epsilon_4\right)+\gamma_4 \epsilon_1=0,\\
%%%%%%%%%%%%
&\left(\bar{\zeta }_2+\Delta +\zeta_2-\epsilon_4\right)\,\gamma_4 -\left(\gamma_2+\delta_4\right)\,\zeta_4 =0,\\
&\left(\alpha_4+\delta_1+\xi_2\right)\,\gamma_4 -\left(2 \gamma_2+\gamma_3+\delta_4\right)\,\xi_4 =0,\\
&\left(\delta_2+\epsilon_2\right)\,\gamma_4 -\left(2 \gamma_2+\gamma_3+\delta_4\right)\,\epsilon_4 +\gamma_1 \zeta_4=0,\\
%%%%%%%%%%%%
&\left(-\bar{\zeta }_3+\Delta -\zeta_3\right)\,\xi_1 -\alpha_1 \zeta_4+\left(-\delta_1+\xi_2+2 \xi_3\right)\,\zeta_1 +\Lambda -\psi_4-\xi_4 \epsilon_1=0,\\
&\Phi_{1011}-\alpha_1 \epsilon_4+\left(\delta_2+\zeta_1-\epsilon_3\right)\,\xi_1 + \left(\xi_3-\delta_1\right)\,\epsilon_1-\psi_1=0,\\
%%%%%%%%%%%%
&\left(-\bar{\zeta }_3+\Delta +\zeta_2\right)\,\xi_2 -\zeta_4 \left(\alpha_2+\xi_1\right)-\delta_1 \zeta_2+\zeta_2 \xi_3-\psi_8+\xi_4 \left(\zeta_1-\epsilon_2\right)=0,\\
&-\alpha_2 \epsilon_4+\delta_2 \xi_2-\delta_1 \epsilon_2-\Lambda -\psi_3+\xi_1 \left(\zeta_2-\epsilon_4\right)+\xi_4 \epsilon_1=0,\\
%%%%%%%%%%%%
&\left(-\bar{\zeta }_3+\Delta +\zeta_2+\zeta_3\right)\,\xi_3 -\zeta_4 \left(\alpha_3+\xi_1\right)-\delta_1 \zeta_3-\psi_7+\xi_4 \left(\zeta_1-\epsilon_3\right)=0,\\
&\Phi_{1111}-\alpha_3 \epsilon_4-\left(\delta_1+\xi_2\right)\,\epsilon_3 +\left(\delta_2+\epsilon_2\right)\,\xi_3 -\psi_4+\xi_1 \left(\zeta_3-\epsilon_4\right)+\xi_4 \epsilon_1=0,\\
%%%%%%%%%%%%
&\left(-\bar{\zeta }_3+\Delta +2 \zeta_2+\zeta_3-\epsilon_4\right)\,\xi_4 -\left(\alpha_4+\delta_1+\xi_2\right)\,\zeta_4 -\psi_6=0,\\
&\Phi_{1211}- \left(\alpha_4+\delta_1+2 \xi_2+\xi_3\right)\,\epsilon_4+\left(\delta_2+2 \epsilon_2+\epsilon_3\right)\,\xi_4 +\zeta_4 \xi_1-\psi_8=0,\\
%%%%%%%%%%%%
&\left(-\bar{\zeta }_3+\Delta -\zeta_3-\epsilon_4\right)\,\epsilon_1 -\left(\delta_2+\zeta_1-\epsilon_2-2 \epsilon_3\right)\,\zeta_1 =0,\\
%%%%%%%%%%%%
&\left(-\bar{\zeta }_3+\Delta +\zeta_2-\epsilon_4\right)\,\epsilon_2 -\left(\delta_2+\zeta_1-\epsilon_3\right)\,\zeta_2 +\zeta_1 \epsilon_4-\zeta_4 \epsilon_1=0,\\
%%%%%%%%%%%%
&\left(-\bar{\zeta }_3+\Delta +\zeta_2+\zeta_3-\epsilon_4\right)\,\epsilon_3 - \left(\delta_2+\zeta_1\right)\,\zeta_3+\zeta_1 \epsilon_4-\zeta_4 \epsilon_1=0,\\
%%%%%%%%%%%%
&\left(-\bar{\zeta }_3+\Delta +2 \zeta_2+\zeta_3-\epsilon_4\right)\,\epsilon_4 - \left(\delta_2+\zeta_1+\epsilon_2\right)\,\zeta_4=0.
\end{align}

\section{The full expression of the Bianchi identities}
\label{app:bian}
In the following Subsection we write the full expression of all the sets of Bianchi identities in turn.

\subsection{The first set of Bianchi identities}

The first of the Bianchi identities is given by:
\begin{align}
D\,\psi_2-\delta_1\,\psi_0&= \alpha_4 \psi_0-\left(\alpha_2+\alpha_3\right) \psi_2+\alpha_1 \psi_5,\\\nonumber
\Delta\,\psi_0-\delta_2\,\psi_1&=\psi_1 \left(-\zeta_1+\epsilon_2-\epsilon_3\right)-\psi_0 \left(\epsilon_4-2 \zeta_3\right)+\frac{1}{3} \Big[\left(-2 \bar{\alpha }_3+\bar{\xi }_1-2 \alpha_1+2 \alpha_2+D\right) \Phi_{2022}+\\
&+2 \left(\delta_2+\epsilon_2-\epsilon_3\right) \Phi_{1011}-2 \epsilon_1 \Phi_{1111}\Big],\\\nonumber
D\,\psi_4-\delta_1\,\psi_1&= -\psi_1 \left(\alpha_4-2 \xi_2\right)+\psi_4 \left(-\alpha_2+\alpha_3-2 \xi_1\right)+\alpha_1 \psi_7+\frac{1}{3} \Big[\bar{\beta }_4 \Phi_{0110}-2 \alpha_1 \Phi_{1211}+\\
&+2 \beta_1 \Phi_{1201}+\left(\delta_1-2\xi_3\right) \Phi_{1011}+\left(\delta_2+2 \epsilon_2\right) \Phi_{0100}-2 \epsilon_4 \Phi_{0000}\Big],\\\nonumber
\Delta\,\psi_1-\delta_2\,\psi_3&= \zeta_4 \psi_0+\psi_1 \left(-\zeta_2+\zeta_3-2 \epsilon_4\right)+\psi_3 \left(2 \epsilon_2-\zeta_1\right)+\\
&+\frac{1}{3} \Big[\left(\delta_1-2 \xi_1+2 \xi_2\right) \Phi_{2022}-2 \epsilon_4 \Phi_{1011}\Big],\\\nonumber
D\,\psi_5-\delta_1\,\psi_2&=\psi_2 \left(-2 \alpha_4+\xi_2-\xi_3\right)+\psi_5 \left(2 \alpha_3-\xi_1\right)+\psi_0 \epsilon_4+\frac{1}{3} \Big[\alpha_4 \Phi_{1011}+\alpha_1 \Phi_{1211}+\\\nonumber
&+2 \Big[\left(-2 \bar{\alpha }_3+\bar{\xi }_1-2 \alpha_3+D\right) \Phi_{1111}+\left(2 \beta_3-\delta_3\right) \Phi_{1101}-\beta_1 \Phi_{1201}+\\
&-\gamma_1 \Phi_{0110}+\zeta_1 \Phi_{0100}\Big]+\left(-2 \bar{\zeta }_3+\bar{\epsilon }_4+\Delta -2 \zeta_3\right) \Phi_{0000}+\left(2 \gamma_3-\delta_4\right) \Phi_{0010}\Big],\\\nonumber
\Delta\,\psi_5-\delta_2\,\psi_7&=-\psi_5 \left(\epsilon_4-2 \zeta_3\right)+\psi_7 \left(-2 \zeta_1+\epsilon_2-\epsilon_3\right)+\psi_6 \epsilon_1+\\\nonumber
&+\frac{1}{3} \Big[\left(-2 \bar{\alpha }_3+\bar{\xi }_1-2 \alpha_3+D\right) \Phi_{2222}+2 \left(-2 \bar{\zeta}_3+\bar{\epsilon }_4+\Delta -2 \zeta_3\right) \Phi_{1111}+\\
&+2 \alpha_4 \Phi_{2122}+\left(2 \beta_3-\delta_3\right) \Phi_{2212}+2 \left(2 \gamma_3-\delta_4\right) \Phi_{1121}+2 \zeta_4 \Phi_{1011}+2 \zeta_1 \Phi_{1211}\Big],\\\nonumber
\Delta\,\psi_2-\delta_2\,\psi_4&=\psi_2 \left(\zeta_3-\epsilon_4\right)+\psi_4 \left(\epsilon_2-2 \zeta_1\right)+\psi_8 \epsilon_1+\frac{1}{3} \Big[2 \Big(\bar{\beta }_4 \Phi_{1121}-\alpha_1 \Phi_{2222}+\beta_1 \Phi_{2212}\Big)+\\\nonumber
&+\left(-2 \bar{\alpha }_3+\bar{\xi }_1+2 \alpha_2+D\right) \Phi_{2122}+\left(-2 \bar{\zeta}_3+\bar{\epsilon }_4+\Delta -2 \zeta_3+2 \epsilon_4\right) \Phi_{1011}+\\
&+\left(\delta_2-2 \epsilon_3\right) \Phi_{1111}\Big],\\\nonumber
   %%%%%%%%%%%%%%%%%%%%%%555
D\,\psi_8-\delta_1\,\psi_3&=\psi_3 \left(\xi_2-\alpha_4\right)+\psi_8 \left(\alpha_3-\xi_1\right)+\frac{1}{3} \Big[\left(\delta_2+2 \epsilon_2\right) \Phi_{0200}-2 \xi_1 \Phi_{1211}+\\
&-2 \epsilon_4 \left(\Phi_{0100}+\Phi_{1011}\right)\Big]\\\nonumber
%%%%%%%%%%%%%%%%%%%%%%%%%%%%%%%%55
D \,\psi_6 -\delta_1\,\psi_8&= -\psi_8 \left(\alpha_4+\xi_3\right)+\alpha_3 \psi_6+\psi_3 \epsilon_4+\frac{1}{3} \Big[4 \bar{\gamma }_1 \Phi_{1201}-2 \zeta_4 \Phi_{0100}-2 \epsilon_4 \Phi_{1111}+\\
&+\left(-2 \bar{\zeta }_3+\bar{\epsilon }_4+\Delta +2 \zeta_2\right) \Phi_{0200}+2 \gamma_4 \Phi_{0110}+2 \left(\delta_1+\xi_2-\xi_3\right) \Phi_{1211}\Big],\\\nonumber
\Delta\,\psi_7-\delta_2\,\psi_6&=\psi_6 \left(-\zeta_1+2 \epsilon_2+\epsilon_4\right)-\psi_7 \left(\zeta_2-\zeta_3+\zeta_4+\epsilon_4\right)+\frac{1}{3} \Big[\bar{\gamma }_1 \Phi_{2212}+2 \epsilon_4 \Phi_{2122}+\\
&+\left(-3 \bar{\zeta }_3+\bar{\epsilon }_4+\Delta +2 \zeta_2\right) \Phi_{1211}+2 \gamma_4 \Phi_{1121}+\left(\delta_1-2 \xi_3\right) \Phi_{2222}-2 \zeta_4 \Phi_{1111}\Big],\\\nonumber
D\,\psi_7-\delta_1\,\psi_4&=\psi_4 \left(-2 \alpha_4+\xi_2-\xi_3\right)+\psi_7 \left(2 \alpha_3-\xi_1\right)+\psi_1 \epsilon_4+\frac{1}{3} \Big[2 \left(\bar{\gamma }_1+\beta_4\right) \Phi_{1101}+\\\nonumber
&+\left(-2 \bar{\alpha }_3+\bar{\xi }_1+2 \alpha_2+D\right) \Phi_{1211}+\left(-2 \bar{\zeta }_3+\bar{\epsilon }_4+\Delta -2 \zeta_3\right) \Phi_{0100}+2 \epsilon_4 \Phi_{1011}+\\
&+\left(-2 \alpha_4+\delta_1-2 \xi_3\right) \Phi_{1111}-\left(2 \beta_2+\delta_3\right) \Phi_{1201}+\left(2 \gamma_3-\delta_4\right) \Phi_{0110}+2 \zeta_1 \Phi_{0200}\Big],\\\nonumber
\Delta\,\psi_4-\delta_2\,\psi_8&=\psi_4 \left(\zeta_3-\epsilon_4\right)+\psi_8 \left(\epsilon_2-\zeta_1\right)+\frac{1}{3} \Big[\left(\delta_1-2 \xi_3\right) \Phi_{2022}+\left(\delta_2+2 \epsilon_2\right) \Phi_{1211}+\\
&+2 \epsilon_4 \left(\Phi_{2022}-\Phi_{1111}\right)\Big],\\
\delta_4\,\psi_0&=2 \gamma_3 \psi_0-\frac{2}{3} \left(\bar{\beta }_4 \Phi_{1011}+\epsilon_1 \Phi_{1101}\right),\\
\delta_4\,\psi_1&=\gamma_4 \psi_0+\gamma_3 \psi_1-\frac{2}{3} \epsilon_1 \Phi_{1201},\\\nonumber
\delta_3\,\psi_2&=\beta_3 \psi_2+\beta_1 \psi_5+\frac{1}{3} \Big[\left(-\bar{\alpha }_3+2 \bar{\xi }_1-2 \alpha_3+D\right) \Phi_{1211}+2 \zeta_1 \Phi_{0110}+\\
&+\left(2 \bar{\alpha }_2-\bar{\alpha}_3+2 \bar{\xi }_1-2 \alpha_3+D\right) \Phi_{1121}+\left(\bar{\zeta }_2-\bar{\zeta }_3+\bar{\epsilon }_4+\Delta -2 \zeta_3\right) \Phi_{0010}\Big],\\
\delta_4\,\psi_2&=2\gamma_3 \psi_2+\frac{1}{3} \Big[-\bar{\beta }_4 \Phi_{1111}-2 \alpha_1 \Phi_{2212}+2 \beta_1 \Phi_{2202}+\left(\delta_2-2 \epsilon_3\right) \Phi_{1101}\Big],\\
\delta_3\,\psi_4&=\beta_3 \psi_4+\beta_1 \psi_7+\frac{1}{3} \Big[-\bar{\gamma }_1 \Phi_{1011}+\left(\delta_2-2 \epsilon_3\right) \Phi_{0110}-2 \xi_1 \Phi_{1121}\Big],\\
\delta_4\,\psi_4&=\gamma_4 \psi_2+\gamma_3 \psi_4+\frac{1}{3} \left(\left(\bar{\gamma }_2+\delta_2-2 \epsilon_3\right) \Phi_{1201}-2 \xi_1 \Phi_{2212}\right),\\\nonumber
\delta_3\,\psi_5&=2 \beta_3 \psi_5+\frac{1}{3} \Big[\left(-\bar{\alpha }_3+2 \bar{\xi }_1-2 \alpha_3+D\right) \Phi_{1211}+2 \zeta_1 \Phi_{0110}+\\
&+\left(2 \bar{\alpha }_2-\bar{\alpha}_3+2 \bar{\xi }_1-2 \alpha_3+D\right) \Phi_{1121}+\left(\bar{\zeta }_2-\bar{\zeta }_3+\bar{\epsilon }_4+\Delta -2 \zeta_3\right) \Phi_{0010}\Big],\\\nonumber
\delta_4\,\psi_5&=2 \gamma_3 \psi_5+\frac{1}{3} \Big[\left(\bar{\alpha }_2-\bar{\alpha }_3+\bar{\xi }_1-2 \alpha_3+D\right) \Phi_{2212}+2 \zeta_1 \Phi_{1201}+\\
&+2 \left(\bar{\zeta}_2-\bar{\zeta }_3+2 \bar{\epsilon }_4+\Delta -2 \zeta_3\right) \Phi_{1101}+\left(2 \beta_3-\delta_3\right) \Phi_{2202}\Big],\\
\delta_3\,\psi_6&=2 \beta_3 \psi_6-\frac{2}{3} \Big[\left(\bar{\gamma }_1+\epsilon_4\right) \Phi_{1211}+\zeta_4 \Phi_{0110}\Big],\\\nonumber
\delta_3\,\psi_7&=2 \beta_3 \psi_7+\frac{1}{3} \Big[-\bar{\gamma }_1 \Phi_{1111}+\left(\bar{\zeta }_2-\bar{\zeta }_3+\bar{\epsilon }_4+\Delta -2 \zeta_3\right) \Phi_{0110}+\\
&-2 \alpha_4 \Phi_{1211}+\left(\delta_1-2 \xi_3\right) \Phi_{1121}\Big],\\\nonumber
\delta_4\,\psi_7&=\gamma_4 \psi_5+\gamma_3 \psi_7+\frac{1}{3} \Big[\bar{\gamma }_1 \Phi_{2202}+2 \left(\bar{\zeta }_2-\bar{\zeta }_3+2 \bar{\epsilon }_4+\Delta +\zeta_2-\zeta_3\right) \Phi_{1201}+\\
&+\left(\delta_1-2 \xi_3\right) \Phi_{2212}-2 \zeta_4 \Phi_{1101}\Big],\\
\delta_3\,\psi_8&=\beta_1 \psi_6+\beta_3 \psi_8-\frac{2}{3} \epsilon_4 \Phi_{0110}.
\end{align}

\subsection{The second set of Bianchi identities}
The second set of Bianchi identities is  given by:
\begin{align}\nonumber
&3 D \Lambda +\beta_1 \Phi_{1201}+\epsilon_1 \Phi_{0200}+3\left(2 \gamma_1-\bar{\beta }_4\right) \Phi_{0110}+\left(-2 \bar{\alpha }_3+\bar{\xi }_1+2 \alpha_3+D-\xi_1\right) \Phi_{1111}+\\\nonumber
&+\left(-2 \bar{\zeta }_3+\bar{\epsilon }_4+\Delta +2 \zeta_3-\epsilon_4\right) \Phi_{0000}=\left(\alpha_4+\delta_1-\xi_2+\xi_3\right) \Phi_{1011}+\left(2 \beta_3+\delta_3\right) \Phi_{1101}+\\
&+\left(2 \gamma_3+\delta_4\right) \Phi_{0010}+\left(\delta_2+2 \zeta_1-\epsilon_2+\epsilon_3\right) \Phi_{0100},\\\nonumber
\\\nonumber
&3 \Delta  \Lambda+\left(-2 \bar{\alpha }_3+\bar{\xi }_1+2 \alpha_3+D-\xi_1\right) \Phi_{2222}+\left(-2 \bar{\zeta }_3+\bar{\epsilon }_4+\Delta +2 \zeta_3-\epsilon_4\right) \Phi_{1111}+\\\nonumber
&+\left(\epsilon_4-\alpha_4\right) \Phi_{2022}=\left(\alpha_4+\delta_1-\xi_2+\xi_3\right) \Phi_{2122}+\left(2 \beta_3+\delta_3\right) \Phi_{2212}+\zeta_4 \Phi_{1011}+\\
&+\Phi_{1211} \left(2 \gamma_3+\delta_2+\delta_4+\zeta_1-\epsilon_2+\epsilon_3\right),\\\nonumber
\\\nonumber
&3 \delta_1 \Lambda+\left(-2 \bar{\zeta }_3+\bar{\epsilon }_4+\Delta -\zeta_2+\zeta_3-2 \epsilon_4\right) \Phi_{0100}+\alpha_4 \Phi_{1111}+\left(\beta_2-\beta_3-\delta_3\right) \Phi_{1201}+\\
&+\left(\gamma_2-\gamma_3-\delta_4\right) \Phi_{0110}+\zeta_4 \Phi_{0000} =\beta_4 \Phi_{1101}+\gamma_4 \Phi_{0010}+\left(\delta_2+\zeta_1-2 \epsilon_2\right) \Phi_{0200}+\epsilon_4 \Phi_{1011},\\\nonumber
\\\nonumber
&3\delta_2 \Lambda+ \left(-2 \bar{\alpha }_3+\bar{\xi }_1-\alpha_2+\alpha_3-\alpha_4-\delta_1+D-\xi_1+2 \xi_2\right)\Phi_{2022}-\beta_1 \Phi_{2212}-\xi_1 \Phi_{2122}\\
&+\left(-2 \bar{\zeta }_3+\bar{\epsilon }_4+\Delta -\zeta_2+\zeta_3-\epsilon_4\right) \Phi_{1011}+\alpha_1 \Phi_{2222}+\zeta_1 \Phi_{1111} =\left(\gamma_1+\epsilon_1\right) \Phi_{1211},\\\nonumber
\\\nonumber
&3 \delta_3 \Lambda +\bar{\gamma }_1 \Phi_{1011}+\left(\bar{\alpha }_2-\bar{\alpha }_3+2 \bar{\xi }_1+2 \alpha_3+D-\xi_1\right) \Phi_{1121}+\\
&+\left(\bar{\zeta }_2-\bar{\zeta }_3+\bar{\epsilon }_4+\Delta +2 \zeta_3-\epsilon_4\right) \Phi_{0010}=\left(\delta_2+2\zeta_1-\epsilon_2+\epsilon_3\right) \Phi_{0110},\\\nonumber
\\\nonumber
&3 \delta_4 \Lambda +\bar{\beta }_4 \Phi_{1211}+\left(\bar{\alpha }_2-\bar{\alpha }_3+\bar{\xi }_1+2 \alpha_3+D-\xi_1\right) \Phi_{2212}-\left(2 \beta_3+\delta_3\right) \Phi_{2202}+\\
&+\left(\bar{\zeta }_2-\bar{\zeta }_3+2 \bar{\epsilon }_4+\Delta +2 \zeta_3-\epsilon_4\right) \Phi_{1101}=\Phi_{1201}
   \left(\bar{\gamma }_2-\bar{\gamma }_3+\delta_2+\zeta_1-\epsilon_2+\epsilon_3\right).
\end{align}

\subsection{The third  set of Bianchi identities}

The third set of Bianchi identities is given by (using the definition of the Weyl scalar):
\begin{align}
\nabla_{\boldsymbol{a}\dot{\boldsymbol{b}}}\,\Big[ \psi_{\boldsymbol{cdef}}+\dfrac{\Lambda}{3} \, \Big( \boldsymbol{\epsilon}_{\boldsymbol{ce}}\boldsymbol{\epsilon}_{\boldsymbol{df}}+ \boldsymbol{\epsilon}_{\boldsymbol{cf}}\boldsymbol{\epsilon}_{\boldsymbol{de}} \Big) \Big]=0.
\end{align}
Explicitely, we find:
\begin{align}
&\left(-2 \alpha_2+2 \alpha_3+D\right)\psi_0 =-2 \alpha_1 \left(\psi_2-\psi_1\right),\\
&\left(\delta_2-2 \epsilon_2+2 \epsilon_3\right)\psi_0=-2 \left(\psi_2-\psi_1\right) \epsilon_1,\\
&\left(-2 \beta_2+2 \beta_3+\delta_3\right)\psi_0=-2 \beta_1 \left(\psi_2-\psi_1\right),\\
&\left(-3 \alpha_2+\alpha_3+D\right)\psi_1=-\alpha_4 \psi_0-\alpha_1 \left(\psi_3-2 \psi_4\right),\\
&\left(-3 \beta_2+\beta_3+\delta_3\right)\psi_1=\beta_1 \left(\psi_3-2 \psi_4\right)-\beta_4 \psi_0,\\
&\left(\delta_2-\epsilon_2+3 \epsilon_3\right)\psi_2 =-\psi_0 \left(-\epsilon_4\right)-\left(\psi_5-2 \psi_4\right) \epsilon_1,\\
&\left(D-4 \alpha_2\right)\psi_3 +\frac{D \Lambda }{3}=-2 \alpha_4 \psi_1-2 \alpha_1 \psi_8,\\
&\left(\Delta -4 \zeta_2\right)\psi_3 +\frac{\Delta  \Lambda }{3}=-2 \zeta_4 \psi_1-2 \zeta_1 \psi_8,\\
&\delta_3 \left(\frac{\Lambda }{3}+\psi_3\right)=-2 \beta_4 \psi_1+4 \beta_2 \psi_3-2 \beta_1 \psi_8,\\
&\delta_4 \left(\frac{\Lambda }{3}+\psi_3\right)=-2 \gamma_4 \psi_1+4 \gamma_2 \psi_3-2 \gamma_1 \psi_8,\\
&\delta_1 \left(\frac{\Lambda }{3}+\psi_5\right)=-4 \xi_3 \psi_5+2 \xi_1 \psi_7+2 \psi_2 \epsilon_4,\\
&\delta_2 \left(\frac{\Lambda }{3}+\psi_5\right)=2 \psi_2 \epsilon_4-4 \psi_5 \epsilon_3+2 \psi_7 \epsilon_1,\\
&\left(\Delta -2 \zeta_2+2 \zeta_3\right)\psi_6 =-2 \zeta_4 \left(\psi_7-\psi_8\right),\\
&\left(\delta_1-2 \xi_2+2 \xi_3\right)\psi_6 =-2 \left(\psi_7-\psi_8\right) \epsilon_4,\\
&\left(-2 \gamma_2+2 \gamma_3+\delta_4\right)\psi_6 =-2 \gamma_4 \left(\psi_7-\psi_8\right),\\
&\left(\delta_1-\xi_2+3 \xi_3\right)\psi_7 =\xi_1 \psi_6-\left(\psi_5-2 \psi_4\right) \epsilon_4,\\
&\left(\delta_2-\epsilon_2+3 \epsilon_3\right)\psi_7 =-\left(\psi_5-2 \psi_4\right) \epsilon_4+\psi_6 \epsilon_1,\\
&\left(\Delta -3 \zeta_2+\zeta_3\right)\psi_8 =\zeta_4 \left(\psi_3-2 \psi_4\right).\zeta_1 \psi_6,\\
&\left(-3 \beta_2+\beta_3+\delta_3\right)\psi_8 =\beta_4 \left(\psi_3-2 \psi_4\right)-\beta_1 \psi_6.
\end{align}

\subsection{The last set of Bianchi identities}
The last set of Bianchi identities is:
\begin{equation}
\nabla_{\boldsymbol{a}\dot{\boldsymbol{b}}}\phi_{\boldsymbol{cd}\dot{\boldsymbol{e}}\dot{\boldsymbol{f}}}=0.
\end{equation}

\begin{align}
&\left(2 \bar{\alpha }_3+2 \alpha_3+D\right) \Phi_{0000}=2 \alpha_1 \Phi_{0100},\\
&\left(\delta_1+2 \xi_3\right) \Phi_{0000}=2 \xi_1 \Phi_{0100},\\
&\left(2 \beta_3+\delta_3\right) \Phi_{0000}=2 \beta_1 \Phi_{0100},\\
&\left(2 \bar{\alpha }_3-\alpha_2+\alpha_3+D\right) \Phi_{0100}=-\alpha_4 \Phi_{0000}+\alpha_1 \Phi_{0200},\\
&\left(-\beta_2+\beta_3+\delta_3\right) \Phi_{0100}=-\beta_4 \Phi_{0000}+\beta_1 \Phi_{0200},\\
&\left(2 \bar{\alpha }_3-2 \alpha_2+D\right) \Phi_{0200}=-2 \alpha_4 \Phi_{0100},\\
&\left(\delta_3-2 \beta_2\right) \Phi_{0200}=-2 \beta_4 \Phi_{0100},\\
&\left(-\bar{\alpha }_2+\bar{\alpha }_3+2 \alpha_3+D\right) \Phi_{0010}=2 \alpha_1 \Phi_{0110},\\
&\left(\delta_1+2 \xi_3\right) \Phi_{0010}=2 \xi_1 \Phi_{0110},\\
&\left(-2 \bar{\alpha }_2+2 \alpha_3+D\right) \Phi_{0020}=0,\\
&\left(\delta_1+2 \xi_3\right) \Phi_{0020}=0,\\
&\left(-\bar{\alpha }_2+\bar{\alpha }_3-\alpha_2+\alpha_3+D\right) \Phi_{0110}=-alpha_4 \Phi_{0010},\\
&\left(-\beta_2+\beta_3+\delta_3\right) \Phi_{0110}=-\beta_4 \Phi_{0010},\\
&\left(\delta_1-\xi_2+\xi_3\right) \Phi_{0110}=-\epsilon_4 \Phi_{0010},\\
&\left(\delta_1-\xi_2+\xi_3\right) \Phi_{1011}=-\xi_1 \Phi_{1111},\\
&\left(\delta_1+2 \xi_3\right) \Phi_{1101}=\xi_1 \Phi_{1201},\\
&\left(\delta_1+\xi_1-2 \xi_2\right) \Phi_{2022}=-\xi_1 \Phi_{2122},\\
&\left(\delta_2+\epsilon_1-2 \epsilon_2\right) \Phi_{2022}=-\epsilon_1 \Phi_{2122},\\
&\left(\delta_1+2 \xi_3\right) \Phi_{2202}=0,\\
&\left(\delta_2+2 \epsilon_3\right) \Phi_{2202}=0,\\
&\left(\delta_1-\xi_2+\xi_3\right) \Phi_{1201}=0,\\
&\left(-\beta_2+\beta_3+\delta_3\right) \Phi_{1201}=0,\\
&\left(2 \bar{\zeta }_3+\Delta +2 \zeta_3\right) \Phi_{2222}=\zeta_4 \left(\Phi_{2022}+\Phi_{2122}\right),\\
&\left(2 \gamma_3+\delta_4\right) \Phi_{2222}=\gamma_4 \left(\Phi_{2022}+\Phi_{2122}\right),\\
&\left(-\gamma_2+\gamma_3+\delta_4\right) \Phi_{1211}=-\gamma_4 \Phi_{1111},\\
&\left(2 \gamma_3+\delta_4\right) \Phi_{1211}=0,\\
&\left(\delta_2+2 \epsilon_3\right) \Phi_{1211}=0,\\
&\left(-\bar{\zeta }_2+\bar{\zeta }_3+\Delta +2 \zeta_3\right) \Phi_{2212}=0,\\
&\left(\delta_2+2 \epsilon_3\right) \Phi_{2212}=0.
\end{align}

%%%%%%%%%%%%%%%%%%%%%%%%%%%%%%%%%%%%%%%%%%%%%%%%%%%%%
\let\doi\relax

\bibliographystyle{abbrv}
\bibliography{biblio}

@BOOK{EC,
       author = {{Cartan}, Elie},
        title = "{The theory of spinors}",
         year = 1981,
       publisher={Dover publications},
       ISBN={9780486640709}
}

@BOOK{RP,
       author = {{Penrose}, Roger and {Rindler}, Wolfgang},
        title = "{Spinors and spacetime -- Vol. 1}",
         year = 1984,
       publisher={Cambridge University Press},
       doi={10.1017/CBO9780511564048},
       ISBN={9780511564048}
}

@BOOK{RP2,
       author = {{Penrose}, Roger and {Rindler}, Wolfgang},
        title = "{Spinors and spacetime -- Vol. 2}",
         year = 1984,
       publisher={Cambridge University Press},
       doi={10.1017/CBO9780511524486},
       ISBN={9780511524486}
}

@BOOK{sup,
       author = {{Freedman}, Daniel Z. and {van Proeyen}, Antoine},
        title = "{Supergravity}",
         year = 2012,
       publisher={Cambridge University Press},
       doi={10.1017/CBO9781139026833},
       ISBN={9781139026833}
}

@BOOK{SW1,
       author = {{Weinberg}, Steven},
        title = "{The Quantum Theory of Fields - Volume 1 - Foundations}",
         year = 1995,
       publisher={Cambridge University Press},
       doi={10.1017/CBO9781139644167},
       ISBN={9781139644167}
}

@BOOK{SW2,
       author = {{Weinberg}, Steven},
        title = "{The Quantum Theory of Fields - Volume 2 - Modern applications}",
         year = 1996,
       publisher={Cambridge University Press},
       doi={10.1017/CBO9781139644174},
       ISBN={9781139644174}
}

@BOOK{stephani,
       author = {{Stephani}, H. and {Kramer}, D. and {MacCallum}, M. and {Hoenselaers}, C. and {Herlt}, E.},
        title = "{Exact Solutions of Einstein's Field Equations}",
         year = 2003,
       publisher={Cambridge University Press},
       doi={10.1017/CBO9780511535185},
       ISBN={9780511535185}
}

@BOOK{poisson,
       author = {{Poisson}, E.},
        title = "{A Relativist's Toolkit - The Mathematics of Black-Hole Mechanics}",
         year = 2004,
       publisher={Cambridge University Press},
       doi={10.1017/CBO9780511606601},
       ISBN={9780511606601}
}

@unpublished{supc,
  TITLE = {{Un soup{\c c}on de th{\'e}orie des groupes: groupe des rotations et groupe de Poincar{\'e}}},
  AUTHOR = {Delamotte, Bertrand},
  URL = {https://cel.hal.science/cel-00092924},
  NOTE = {Lecture},
  TYPE = {DEA},
  YEAR = {2006},
  MONTH = Sep,
  HAL_ID = {cel-00092924},
  HAL_VERSION = {v1},
}

@ARTICLE{NP,
       author = {{Newman}, Ezra and {Penrose}, Roger},
        title = "{An Approach to Gravitational Radiation by a Method of Spin Coefficients}",
      journal = {Journal of Mathematical Physics},
         year = 1962,
        month = may,
       volume = {3},
       number = {3},
        pages = {566-578},
          doi = {10.1063/1.1724257}
}

@ARTICLE{P1,
       author = {{Penrose}, Roger},
        title = "{A spinor approach to general relativity}",
      journal = {Annals of Physics},
         year = 1960,
        month = jun,
       volume = {10},
       number = {2},
        pages = {171-201},
          doi = {10.1016/0003-4916(60)90021-X}
}

@article{bade,
  title={An introduction to spinors},
  author={Bade, William L and Jehle, Herbert},
  journal={Reviews of Modern Physics},
  volume={25},
  number={3},
  pages={714},
  year={1953},
  publisher={APS},
  doi={10.1103/RevModPhys.25.714}
}

@article{witten,
  title={Invariants of general relativity and the classification of spaces},
  author={Witten, Louis},
  journal={Physical Review},
  volume={113},
  number={1},
  pages={357},
  year={1959},
  publisher={APS},
  doi={10.1103/PhysRev.113.357}
}

@article{sp1,
     author = {Cartan, E.},
     title = {Les groupes projectifs qui ne laissent invariante aucune multiplicit\'e plane},
     journal = {Bulletin de la Soci\'et\'e Math\'ematique de France},
     pages = {53--96},
     publisher = {Soci\'et\'e math\'ematique de France},
     volume = {41},
     year = {1913},
     doi = {10.24033/bsmf.916},
     language = {fr},
     url = {https://www.numdam.org/articles/10.24033/bsmf.916/}
}

@ARTICLE{teu1,
       author = {{Teukolsky}, Saul A.},
        title = "{Perturbations of a Rotating Black Hole. I. Fundamental Equations for Gravitational, Electromagnetic, and Neutrino-Field Perturbations}",
      journal = {Ap. J.},
         year = 1973,
        month = oct,
       volume = {185},
        pages = {635-648},
          doi = {10.1086/152444}
}

@ARTICLE{teu2,
       author = {{Press}, William H. and {Teukolsky}, Saul A.},
        title = "{Perturbations of a Rotating Black Hole. II. Dynamical Stability of the Kerr Metric}",
      journal = {Ap. J},
         year = 1973,
        month = oct,
       volume = {185},
        pages = {649-674},
          doi = {10.1086/152445}
}

@ARTICLE{lrr,
       author = {{Sasaki}, M. and {Tagoshi}, H.},
        title = "{Analytic Black Hole Perturbation Approach to Gravitational Radiation}",
      journal = {Living Review in Relativity},
         year = 2003,
       volume = {6},
        pages = {6},
          doi = {10.12942/lrr-2003-6}
}

@article{kono,
  title = {Quasinormal modes of massive fermions in Kerr spacetime: Long-lived modes and the fine structure},
  author = {{Konoplya}, Roman A. and {Zhidenko}, Alexander},
  journal = {Phys. Rev. D},
  volume = {97},
  issue = {8},
  pages = {084034},
  numpages = {10},
  year = {2018},
  month = {Apr},
  publisher = {American Physical Society},
  doi = {10.1103/PhysRevD.97.084034},
  url = {https://link.aps.org/doi/10.1103/PhysRevD.97.084034}
}

@article{hd,
  title = {Quasinormal modes of a d-dimensional regular black hole featuring an integrable singularity},
  author = {{Dong}, Z. and {Zhang}, D. and {Fu}, G. and {Wu}, J.},
  journal = {Eur. Phys. J. C},
  volume = {85},
  pages = {215},
  numpages = {10},
  year = {2025},
  doi = {10.1140/epjc/s10052-025-13926-3}
}

@article{hdc1,
doi = {10.1088/0264-9381/25/3/033001},
url = {https://dx.doi.org/10.1088/0264-9381/25/3/033001},
year = {2008},
month = {jan},
publisher = {},
volume = {25},
number = {3},
pages = {033001},
author = {Coley, A},
title = {Classification of the Weyl tensor in higher dimensions and applications},
journal = {Classical and Quantum Gravity}
}

@article{hdc2,
doi = {10.1088/0264-9381/21/7/L01},
url = {https://dx.doi.org/10.1088/0264-9381/21/7/L01},
year = {2004},
month = {mar},
publisher = {},
volume = {21},
number = {7},
pages = {L35},
author = {A Coley and R Milson and V Pravda and A Pravdová},
title = {Classification of the Weyl tensor in higher dimensions},
journal = {Classical and Quantum Gravity}
}

@article{hdc3,
author = {Milson, R. and Coley, A. and Pravda, V. and Pravdov\'{a}, A.},
title = {ALIGNMENT AND ALGEBRAICALLY SPECIAL TENSORS IN LORENTZIAN GEOMETRY},
journal = {International Journal of Geometric Methods in Modern Physics},
volume = {02},
number = {01},
pages = {41-61},
year = {2005},
doi = {10.1142/S0219887805000491},
URL = {https://doi.org/10.1142/S0219887805000491},
eprint = {https://doi.org/10.1142/S0219887805000491}
}

@article{hdc4,
doi = {10.1088/0264-9381/21/12/007},
url = {https://dx.doi.org/10.1088/0264-9381/21/12/007},
year = {2004},
month = {may},
publisher = {},
volume = {21},
number = {12},
pages = {2873},
author = {V Pravda and A Pravdová and A Coley and R Milson},
title = {Bianchi identities in higher dimensions},
journal = {Classical and Quantum Gravity}
}

@article{JN,
author="A. I. Janis and E. T. Newman",
title="Structure of Gravitational Sources",
journal="Journal of Mathematical Physics",
ISSN="0022-2488",
publisher="AIP Publishing",
year="1965",
month="06",
volume="6",
number="6",
pages="902-914",
DOI="10.1063/1.1704349",
URL="https://cir.nii.ac.jp/crid/1363107369228328704"
}

@article{huges,
  title = {Evolution of circular, nonequatorial orbits of Kerr black holes due to gravitational-wave emission},
  author = {Hughes, Scott A.},
  journal = {Phys. Rev. D},
  volume = {61},
  issue = {8},
  pages = {084004},
  numpages = {28},
  year = {2000},
  month = {Mar},
  publisher = {American Physical Society},
  doi = {10.1103/PhysRevD.61.084004},
  url = {https://link.aps.org/doi/10.1103/PhysRevD.61.084004}
}

@article{aguci,
    author = {Higuchi, Atsushi},
    title = {Symmetric tensor spherical harmonics on the N‐sphere and their application to the de Sitter group SO(N,1)},
    journal = {Journal of Mathematical Physics},
    volume = {28},
    number = {7},
    pages = {1553-1566},
    year = {1987},
    month = {07},
    issn = {0022-2488},
    doi = {10.1063/1.527513},
    url = {https://doi.org/10.1063/1.527513},
    eprint = {https://pubs.aip.org/aip/jmp/article-pdf/28/7/1553/19127081/1553\_1\_online.pdf},
}

@article{leaver,
author = {Leaver, E. W.  and Chandrasekhar, Subrahmanyan },
title = {An analytic representation for the quasi-normal modes of Kerr black holes},
journal = {Proceedings of the Royal Society of London. A. Mathematical and Physical Sciences},
volume = {402},
number = {1823},
pages = {285-298},
year = {1985},
doi = {10.1098/rspa.1985.0119},
URL = {https://royalsocietypublishing.org/doi/abs/10.1098/rspa.1985.0119},
eprint = {https://royalsocietypublishing.org/doi/pdf/10.1098/rspa.1985.0119}
}

@ARTICLE{highKerr1,
       author = {{Kao}, Hsien-Chung and {Tomino}, Dan},
        title = "{Quasinormal modes of Kerr black holes in four and higher dimensions}",
      journal = {\prd},
         year = 2008,
        month = jun,
       volume = {77},
       number = {12},
          eid = {127503},
        pages = {127503},
          doi = {10.1103/PhysRevD.77.127503},
archivePrefix = {arXiv},
       eprint = {0801.4195},
 primaryClass = {gr-qc},
       adsurl = {https://ui.adsabs.harvard.edu/abs/2008PhRvD..77l7503K}
}

@ARTICLE{highKerr2,
       author = {{Lu}, Kai-Peng and {Li}, Wenbin and {Huang}, Jia-Hui},
        title = "{Quasinormal modes and stability of higher dimensional rotating black holes under massive scalar perturbations}",
      journal = {Physics Letters B},
         year = 2023,
        month = oct,
       volume = {845},
          eid = {138147},
        pages = {138147},
          doi = {10.1016/j.physletb.2023.138147},
archivePrefix = {arXiv},
       eprint = {2307.02338},
 primaryClass = {gr-qc},
       adsurl = {https://ui.adsabs.harvard.edu/abs/2023PhLB..84538147L}
}

@article{highKerr3,
doi = {10.1088/1402-4896/ad5eca},
url = {https://doi.org/10.1088/1402-4896/ad5eca},
year = {2024},
month = {jul},
publisher = {IOP Publishing},
volume = {99},
number = {8},
pages = {085023},
author = {Li, Wenbin and Lu, Kai-Peng and LiMing, W and Huang, Jia-Hui},
title = {Five-dimensional Myers-Perry black holes under massive scalar perturbation: bound states and quasinormal modes},
journal = {Physica Scripta}
}

@article{highKerr4,
  title={Constraining extra dimensions using observations of black hole quasi-normal modes},
  author={Mishra, Akash K and Ghosh, Abhirup and Chakraborty, Sumanta},
  journal={The European Physical Journal C},
  volume={82},
  number={9},
  pages={820},
  year={2022},
  publisher={Springer},
  doi={10.1140/epjc/s10052-022-10788-x}
}

@ARTICLE{ref1,
       author = {{Khan}, Saeed Ullah and {Ren}, Jingli},
        title = "{Shadow cast by a rotating charged black hole in quintessential dark energy}",
      journal = {Physics of the Dark Universe},
         year = 2020,
        month = dec,
       volume = {30},
          eid = {100644},
        pages = {100644},
          doi = {10.1016/j.dark.2020.100644},
       adsurl = {https://ui.adsabs.harvard.edu/abs/2020PDU....3000644K}
}

@article{ref2,
  title={Circular motion and QPOs near black holes in Kalb--Ramond gravity},
  author={Jumaniyozov, Shokhzod and Khan, Saeed Ullah and Rayimbaev, Javlon and Abdujabbarov, Ahmadjon and Urinbaev, Sharofiddin and Murodov, Sardor},
  journal={The European Physical Journal C},
  volume={84},
  number={9},
  pages={964},
  year={2024},
  publisher={Springer},
  doi={10.1140/epjc/s10052-024-13351-y}
}

@ARTICLE{ref3,
       author = {{Khan}, Saeed Ullah and {Rayimbaev}, Javlon and {Turaev}, Yunus and {Sirajiddin}, Otaboyev and {Usanov}, Sulton and {Wang}, Weiwei},
        title = "{Energy efficiency and particle dynamics around magnetized black holes with parabolic configuration in STVG}",
      journal = {European Physical Journal C},
         year = 2026,
        month = mar,
       volume = {86},
       number = {3},
          eid = {207},
        pages = {207},
          doi = {10.1140/epjc/s10052-026-15440-6},
       adsurl = {https://ui.adsabs.harvard.edu/abs/2026EPJC...86..207K}
}
\end{document}